\documentclass[numberedappendix,appendixfloats,iop,apj]{emulateapj}
\usepackage{mathrsfs}
\usepackage{graphicx}
\usepackage{amssymb}
\usepackage{epsf}
\usepackage[title]{appendix}
\usepackage{color}

\newcommand{\Mpc}{{\ensuremath{\rm Mpc}}}

\newcommand{\pri}{\prime}
\newcommand{\tr}{\rm tr}

\newcommand{\vu}{\mathbf u}

\newcommand{\tA}{\tilde {A}}
\newcommand{\ts}{\tilde {s}}
\newcommand{\teta}{\tilde {\eta}}

\newcommand{\bA}{\bar{A}}
\newcommand{\bvu}{\bar{\vu}}
\newcommand{\bs}{\bar{s}}

\newcommand{\vv}{\mathbf v}

\newcommand{\vw}{\mathbf w}
\newcommand{\vx}{\mathbf x}
\newcommand{\vA}{\mathbf A}
\newcommand{\vq}{\mathbf q}
\newcommand{\vp}{\mathbf p}

\newcommand{\ve}{\mathbf e}
\newcommand{\vpsi}{\mathbf \Psi}

\newcommand{\vI}{\mathbf I}

\newcommand{\vR}{\mathbf R}

\newcommand{\mP}{\mathbf {\mathcal P}}

\newcommand{\mH}{\mathcal H}
\newcommand{\mL}{\mathcal L}

\newcommand{\Om}{\Omega_m}

\begin{document}

\title{Kinematic Morphology of Large-scale Structure:  Evolution from 
Potential to Rotational Flow}

\author{
Xin Wang\altaffilmark{1},
Alex Szalay\altaffilmark{1}, 
Miguel A. Arag\'{o}n-Calvo\altaffilmark{1},
Mark C. Neyrinck\altaffilmark{1},
Gregory L. Eyink\altaffilmark{2,1}
}

\altaffiltext{1}{Department of Physics \& Astronomy, Johns Hopkins University, 
Baltimore, MD, US, 21218}
\altaffiltext{2}{Department of Applied Mathematics \& Statistics, Johns Hopkins 
University, Baltimore, MD, US, 21218}

\begin{abstract}
  As an alternative way of describing the cosmological velocity field, we 
  discuss the evolution of rotational invariants constructed 
  from the velocity gradient tensor.
  Compared with the traditional divergence-vorticity decomposition, these 
  invariants, defined as coefficients of characteristic equation of the 
  velocity gradient tensor, enable a complete classification of all possible flow 
  patterns in the dark-matter comoving frame, including both potential and vortical
  flows. We show that this tool, first introduced in turbulence two decades ago, 
  proves to be very useful in understanding the evolution of the cosmic web structure, 
  and in classifying its morphology.
  Before shell-crossing, different categories of potential flow are highly 
  associated with cosmic web structure, because of the coherent evolution of density
  and velocity.  This correspondence is even preserved 
  at some level when vorticity is generated after shell-crossing. 
  The evolution from the potential to vortical flow can be 
  traced continuously by these invariants.
  With the help of this tool, we show that the vorticity is generated in 
  a particular way that is highly correlated with the large-scale structure.
  This includes a distinct spatial distribution and different types of alignment between 
  cosmic web and vorticity direction for various vortical flows.
  Incorporating shell-crossing into closed
  dynamical systems is highly non-trivial, but
  we propose a possible statistical explanation for some of these phenomena 
  relating to the internal structure of the three-dimensional invariants space.
\end{abstract}

\keywords{peculiar velocity, vorticity, large-scale structure, cosmic web}

\maketitle

\section{Introduction}
The peculiar velocity $\vu$, together with the density distribution $\rho$ of dark matter, 
encode abundant information about the evolution of large-scale structure of our Universe.
Besides its statistical manifestation via redshift-space distortions \citep{K87,DP83,H92,CFW94}, 
the line-of-sight component of the peculiar flow of individual galaxies can also be 
obtained from distance indicators such as the Tully-Fisher relation \citep{TF77} between 
galaxy luminosity and rotational velocity, and the transverse component from the proper motion of galaxies \citep{NBD12} by highly accurate astrometric experiments like 
{\it Gaia}\footnote{\it http://www.esa.int/Our\_Activities/Operations/Gaia}.   
Such a dataset of large-scale three-dimensional peculiar velocities
would provide valuable information for our theory of the evolution of dark
matter velocity field.

A virtue of three-dimensional velocity data is the ability to study the 
vorticity contribution $\vw =\nabla \times \vu$ \citep{P80,PS09,HA14}, 
which in the standard cosmology model is negligible on large scale and is mainly 
generated after the shell-crossing 
at later epoch, since any initial rotational component will decay rapidly due to the expansion of the Universe.
And for a long time, the divergence contribution $\theta=\nabla \cdot \vu$, as the only
remaining degree of freedom of $\vu$, dominates most of raw theoretical as well as 
observational investigations in this field. 
However, in the nonlinear regime, the vectorial rotational component might contain valuable morphological information that is lost if only the divergence is studied.
To learn more about the cosmic flow in detail, in this paper we are
interested in the gradient tensor of the velocity field $\vu$, 
\begin{eqnarray}
\label{eqn:Aij}
A_{ij}(\vx, \tau) = \frac{\partial u_i}{\partial x_j}(\vx, \tau)
\end{eqnarray}
where $\vx$ is Eulerian position and $\tau$ the comoving time.

The tensor $A_{ij}$ is closely related with cosmic web structure, 
since it characterizes the velocity variations 
around a mass element as it moves away from a void or toward halo/filamentary/wall
structures.
For irrotational flow, $\vu=-\nabla \psi$ where $\psi$ is the 
velocity potential. At the linear order, $\psi$ is proportional to the 
gravitational potential $\Phi$; therefore, the tensor $A_{ij} \approx \nabla_i 
\nabla_j \Phi$ encodes information about the anisotropic gravitational field, which
eventually leads to the formation of the cosmic web structure.
Given the importance of cosmic web in understanding large-scale structure 
formation \citep{BKP96}, many techniques have been developed 
to classify cosmic web in numerical simulation  \citep{PP09,AC10,S11a,SP11b,BSC10,FR09,HP07a,SCP09,SM05}. 
Similar to the method using the tidal field, or equivalently the Hessian matrix of density, 
one usually considers the eigenvalues $\lambda_i$ of tensor $A_{ij}$. 
Neglecting the subtleties in selecting the criteria, in general, entirely positive/negative eigenvalues correspond to an entirely stretching/compressing region, i.e.\ a void/halo, and a matrix with both positive and negative eigenvalues gives wall or filament structures, 
e.g. \cite{HM12}.
However, besides its implicit coordinate system dependence, once the vorticity
is generated, the anti-symmetric tensor $A_{ij}$ gives complex eigenvalues 
$\lambda_i$, and therefore complicates the discussion of flow morphology in such 
language.

\begin{figure*}[htp]
\begin{center}
\includegraphics[width=0.95\textwidth]{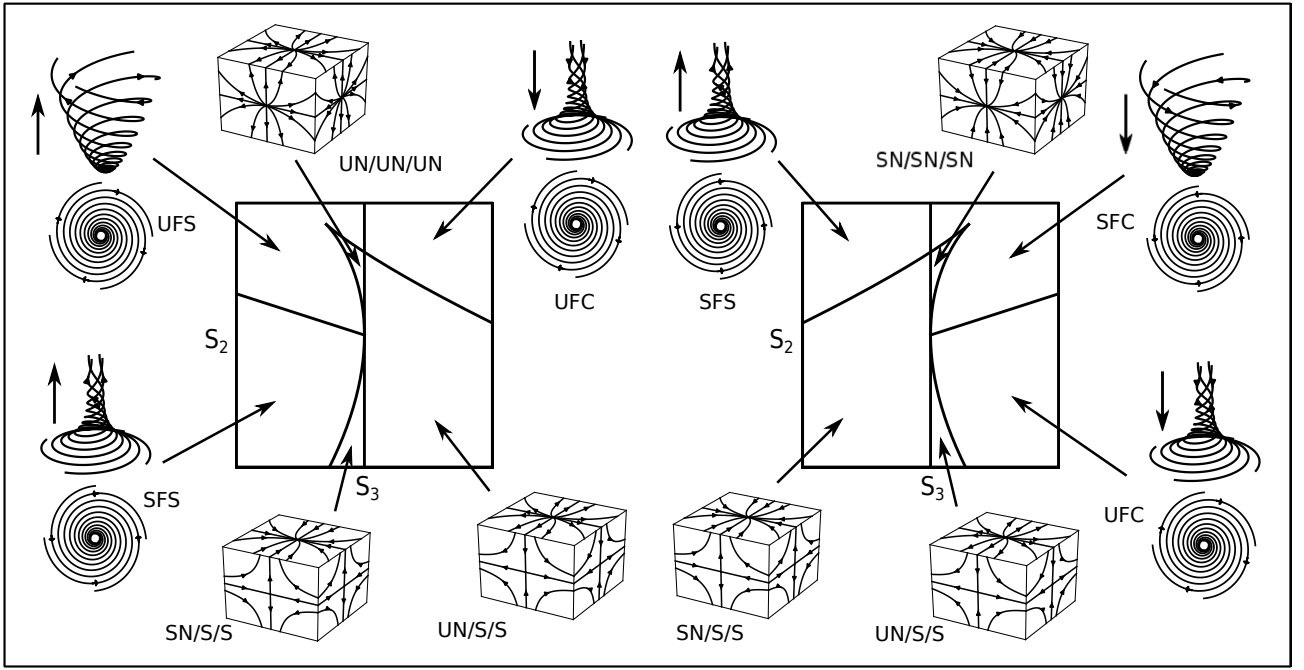}
\end{center}
\caption{ \label{fig:flow_category} Illustration of various cosmic flow types in invariants 
space. We only show 2D $s_3-s_2$ plane with fixed $s_1$; the left panel shows
negative $s_1$, and the right panel shows positive $s_1$. }
\end{figure*}

In this paper, we consider an alternative way of 
describing flow morphologies, as well as its applications.
To eliminate coordinate dependence, one would prefer the scalar fields 
that are rotational invariants built from the velocity
gradient tensor $A_{ij}$. 
The eigenvalue problem of tensor $A_{ij}$ is given by
\begin{eqnarray}
\label{eqn:characteristic_eqa}
\det[ A_{ij} - \lambda \delta_{ij} ] = \lambda^3 + s_1 \lambda^2 + s_2 \lambda + s_3 =0,
\end{eqnarray}
where $\delta_{ij}$ is identity matrix.  The coefficients $s_i$, where $i=\{1,2,3\}$, 
appearing in Eq. (\ref{eqn:characteristic_eqa}), are natural quantities to study.
For various irrotational flows that have been discussed, the criteria in eigenvalue 
space are then mapped into different regions in the invariants space \citep{CPC90}. 
One advantage of working in this parameter space is to avoid  
the complex domain after the emergence of vorticity and $A_{ij}$ becoming anti-symmetric.
It enables one to continuously explore the evolution from potential 
to vortical flow in a uniform framework.

Furthermore, compared with rotation dependent matrix $A_{ij}$, these invariants
are also appropriate device for analytical study of the dynamical evolution of the 
cosmic flow.
Starting from initial fluctuations with negligible rotational degrees of freedom, 
one is able to trace in the invariants space the evolution of kinematic
morphologies of a dark matter element as it travels away from an underdense 
region toward filaments or halos, eventually experiencing various types of 
vorticity in the multi-streaming regime.
However, analytical study of the growth of vorticity is highly non-trivial
\citep{PS09}, 
since it mainly occurs at non-linear small scale which beyond the reach of 
approaches like perturbation theory. 
And the efforts to incorporate the shell-crossing would result in a hierarchy 
of fluid dynamical equations that would need to be closed.
One approach is to model or measure the velocity dispersion and truncate the 
hierarchy \citep[e.g.\ ][]{PS09}.
In this paper, as suggested by our numerical measurement of vortical morphologies,
the stochasticity of the process may provide some new insight.

Even for the collapsed objects like halos,  the full information of tensor $A_{ij}$ is  
important. Theoretically the angular momentum 
of halo is largely explained by the tidal torque theory \citep{H49,P69,D70,W84,WP85},
which explains the formation of halo spin from laminar flow by the misalignment 
between shear and inertia, and it predicts a linear growth of angular momentum. 
But this mechanism is only effective prior to turn-around in the spherical-collapse picture; after this, the collapse dramatically reduces the lever arm
\citep{S09}, and the alignment between halo spin and vorticity direction 
found by \cite{L13a} suggests a separate phase of the growth of halo angular 
momentum.
The coevolution picture between halo spin and vorticity supports our finding
from invariants, because investigations using $N$-body simulations also indicate 
the existence of preferential orientation between halo spin and filaments
\citep{HP07a,SCP09,ZY09,LP13,DP14,ACY14}.  
Indeed, several authors have found mass-dependent alignment of halo 
spins with filaments; low-mass halos tend to be aligned with filaments, and 
high-mass halos have spin perpendicular to the filaments 
\citep{AC07,HC07b,PSP08,SCP09,CP12,AC13}.  
In fact, \citet{TL13} have found evidence for this alignment in SDSS galaxies.

The purposes of this paper are severalfold.  After introducing the invariants
of the velocity gradient tensor at the beginning of section II., we show its ability 
to classify flow morphologies for both potential and vortical flow, and then 
discuss the connection to the cosmic web structure.
Then we try to investigate the dynamical evolution of these invariants, first for
irrotational flow in section III., and then in section IV. study the emergence of vorticity through shell-crossing. 
For simplicity, after deriving the full dynamical equations before shell-crossing,
we first study the Zel'dovich approximation, obtaining the exact numerical solution 
as well as the probability distribution of invariants.
In section IV., we first present the result relating to the vorticity 
measured from N-body simulation, 
and then setup both the dynamical and statistical view for physical interpretation. 
Finally, we discuss and conclude in section V.

Throughout the paper, we make our numerical measurement with two different sets
of $N$-body simulations.
For the purpose of irrotational flow only, we use a simulation with box size 
$100h^{-1}\Mpc$ and $512^3$ particles, hereafter we will denote as `$Sing100$'. 
However the spatial resolution is not high enough to study vortical 
flow which are generated at very small scale. To achieve a clean signal with high
signal-to-noise ratio and meanwhile explore the relation to large-scale structure, 
we use the MIP ({\it Multum In Parvo}, `many things in the same place') ensemble simulation \citep{ACM12}, which is a suite of $N$-body simulations constrained to have the same large-scale structure for initial Fourier modes with wavelengths over $4h^{-1}\Mpc$, but independent small scale realizations.
The combining of many realizations enables one to visually inspect 
the correspondence between rotational flow and cosmic-web.
The ensemble has $256$ realizations \footnote{where only $64$ realizations 
are analyzed in this paper} of a $32 h^{-1} \Mpc$ box with $256^3$
particles each. This is equivalent in terms of effective volume and number of 
particles to a box of $193 ~h^{-1} \Mpc$ of side with $\sim 1540^3$ particles 
containing $\sim 5 \times 10^6$ haloes with a minimum mass of $3.25 \times
10^9 h^{-1} M_{\odot}$.
To numerically estimate the spatial gradient $A_{ij}$, we first construct the velocity 
field on a regular Eulerian grid with Delaunay tessellation \citep{BV96,PS09,SvdW00,PSvdW03}, 
and then take the gradient in real space. 
Although this method is known to have a better noise property in measuring
velocity divergence and vorticity \citep{PS09}, 
it still implicitly involves coarse-graining and is less accurate than more 
sophisticated method with explicit phase space projection \citep{HA14}.

\section{Invariants of Velocity Gradient Tensor and Local Flow Morphology}
The invariants of the velocity gradient tensor have been studied in turbulence 
for more than two decades since their introduction by \cite{CPC90}.  They describe
fundamental and intrinsic properties of small-scale motions in turbulence 
\citep{MC11}.
In this section, we will first present the definition of the invariants
of the velocity gradient tensor, discuss their ability to classify various potential 
and rotational flow morphologies, and then show the connection with cosmic web
structure.

\begin{figure*}[htp]
\begin{center}
\includegraphics[width=1\textwidth]{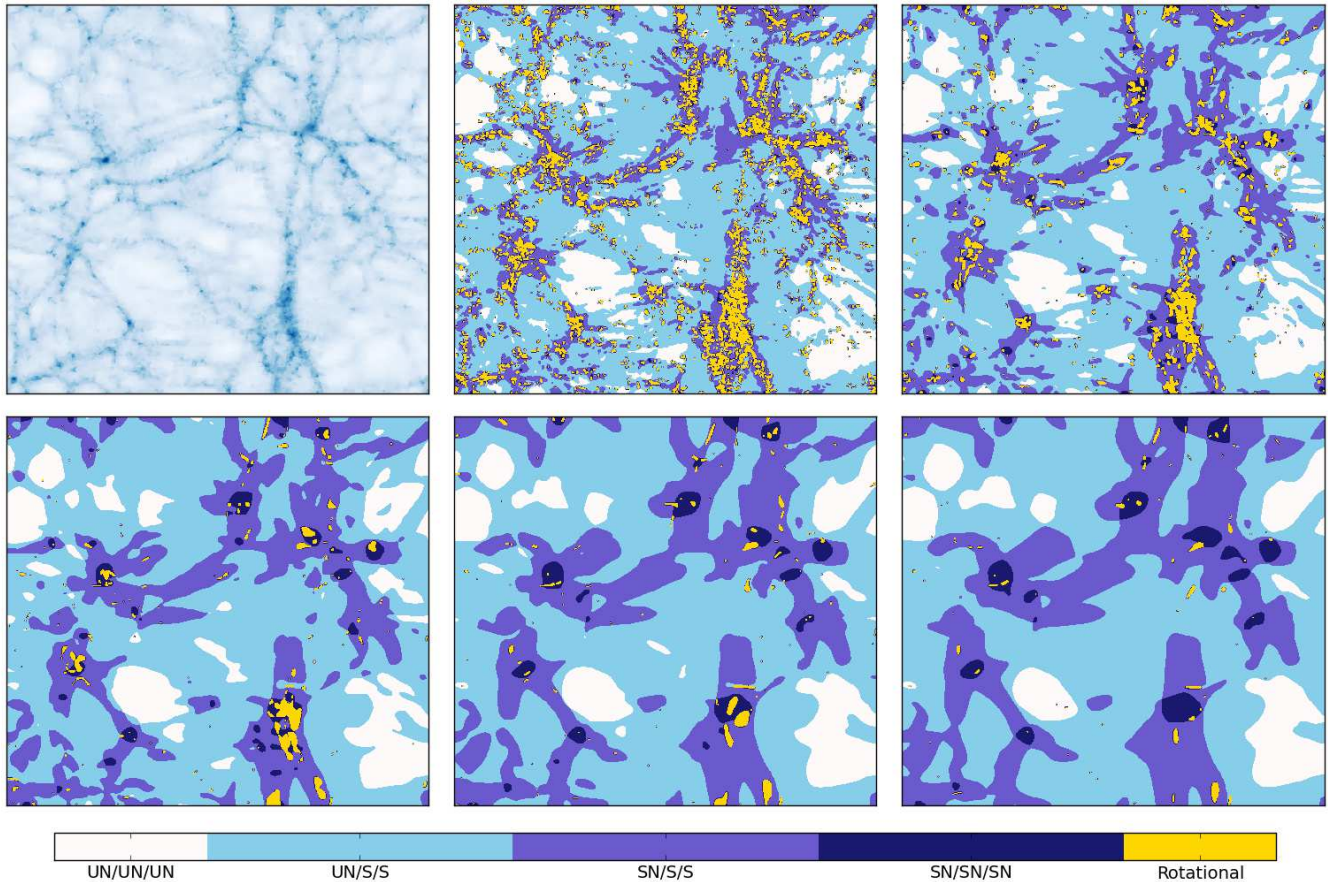}
\end{center}
\caption{ \label{fig:classification_potential} 
Various classifications of cosmic 
flow with different degrees of smoothing, compared with the un-smoothed density
field ({\it first panel}) of the same snapshot of simulation `$Sing100$' at 
redshift $z=0$. The box size is $100\Mpc/h$. 
Various categories are labeled as their kinematical morphologies, the definitions are 
discussed in Sec. 2.2 (also see Fig.\ \ref{fig:flow_category}). 
Starting from the second panel, we use Gaussian 
smoothing scales $R=0.2\Mpc/h, ~ 0.4\Mpc/h, ~ 0.8\Mpc/h, ~ 1.2\Mpc/h$ and $1.5\Mpc/h$.
The rotational region is suppressed increasingly for larger smoothing length. 
}
\end{figure*}

\subsection{Definition of Invariants}
Among all nine elements in tensor $A_{ij}$, six degrees of freedom are 
rotationally invariant, and the other three encode coordinate information.
As discussed in the introduction, to extract the morphological information, we are
particularly interested in the coefficients of the characteristic equation
$\det[\vA-\lambda I] = 0 $, i.e. 
\begin{eqnarray}
\label{eqn:chareq}
 \lambda^3 + s_1 \lambda^2 + s_2 \lambda + s_3=0,
\end{eqnarray}
where the invariants $s_1, s_2, s_3$ are defined as \citep{CPC90}
\begin{eqnarray}
\label{eqn:inv_def}
s_1 &=& -\tr[\vA]= -\theta_{ii}, \nonumber \\
s_2&=& \frac{1}{2} \left ( s_1^2 - \tr[\vA^2] \right ) = 
 \frac{1}{2} (s_1^2 - \theta_{ij}\theta_{ji} -\omega_{ij}\omega_{ji} ) \nonumber \\
s_3&=& - \det[\vA] = \frac{1}{3} \left (-s_1^3 + 3s_1 s_2 - \tr[\vA^3] \right) 
\nonumber \\ 
&=& \frac{1}{3} \left( -s_1^3  + 3s_1 s_2 -  \theta_{ij}\theta_{jk}\theta_{ki} - 
3 \omega_{ij} \omega_{jk} \theta_{ki}  \right).
\end{eqnarray}
Here we have decomposed $A_{ij}$ into a symmetric part $\theta_{ij}$ and an anti-symmetric 
part $\omega_{ij}$,
\begin{eqnarray}
A_{ij} = A_{(ij)}  + A_{[ij]} =  \theta_{ij} + \omega_{ij},  
\end{eqnarray}
where 
\begin{eqnarray}
\theta_{ij} = \frac{1}{2}(A_{ij} + A_{ji}), \qquad
\omega_{ij} = \frac{1}{2}(A_{ij} - A_{ji}). 
\end{eqnarray}

The tensor $\theta_{ij}$ denotes the rate of deformation of a fluid element and 
$\omega_{ij}$ the rate of rotation, which relates to the vorticity vector 
$\omega_i$ through $\omega_i=-\epsilon_{ijk}\omega_{jk}$, where $\epsilon_{ijk}$ is the
totally anti-symmetric Levi-Civita symbol.
To better understand the physics of these invariants $s_i$,
we could further decompose the rate of deformation into isotropic and anisotropic
part, i.e. the divergence $\theta$ and the shear tensor $\sigma_{ij}$
\begin{eqnarray}
\theta_{ij} = \sigma_{ij} + \frac{1}{3}  \theta  \delta^{K}_{ij},
\end{eqnarray}
where $\delta^K_{ij}$ is the Kronecker delta, 
and also define the magnitude of tensor $\sigma_{ij}$ and $\omega_{ij}$
\footnote{The vorticity amplitude $\omega$ of $\omega_{ij}$ defined here relates to 
the amplitude of vorticity vector via $|\omega_i|^2 = 4 \omega^2 $.} as
\begin{eqnarray}
\sigma =\left (\frac{1}{2}\sigma_{ij} \sigma_{ij} \right)^{1/2},  \qquad
\omega = \left (\frac{1}{2}\omega_{ij} \omega_{ij} \right )^{1/2}.
\end{eqnarray}
With the above definitions, the invariants we just introduced can be expressed as
\begin{eqnarray}
\label{eqn:si_phys}
s_1 &=& -\theta \nonumber \\
s_2 &=&  \frac{1}{3} \theta^2  - \sigma^2 + \omega^2 \nonumber\\
s_3 &= & -\frac{1}{27}\theta^3 + \frac{1}{3}\theta\left(\sigma^2 - \omega^2\right)
 - \frac{1}{3}\sigma_{ij}\sigma_{jk}\sigma_{ki} - \frac{1}{4}\sigma_{ij}\omega_i\omega_j. \nonumber\\
\end{eqnarray}
In turbulence, many studies concentrate on incompressible flow at
$\theta=0$, invariants $s_2=\omega^2-\sigma^2$ and $s_3= -\sigma_{ij} 
( \sigma_{jk}\sigma_{ki} /3 + \omega_{jk}\omega_{ki})$ then describe the balance only
between the shearing $\sigma_{ij}$ and vorticity $\omega_{ij}$.  
It is then reasonable to discuss these two contributions separately, and therefore
one could define remaining scalar degree of freedom of $A_{ij}$ as, e.g. the symmetric 
$s^{(s)}_i$ (or anti-symmetric $s^{(a)}_i$) contribution to $s_2$ and $s_3$, where
\begin{eqnarray}
s_i = s^{(s)}_i + s^{(a)}_i,  \qquad i=(2,3).
\end{eqnarray}
In cosmology, however, $\theta$ is in general nonzero, the anti-symmetric contribution could be expressed as
\begin{eqnarray}
s^{(a)}_{2} &=& - \frac{1}{2} \omega_{ij}\omega_{ji} = \omega^2, \nonumber\\
s^{(a)}_{3} &=& s_1 s^{(a)}_2 - \omega_{ij}\omega_{jk}\theta_{ki} \nonumber \\
 &=& -\frac{1}{3} \theta \omega^2 - \frac{1}{4}\sigma_{ij}\omega_i\omega_j.
\end{eqnarray}
As will be seen later, before shell-crossing, $s^{(a)}_{2}$ and  $s^{(a)}_{3}$ vanish,
eigenvalues $\lambda_i$ of matrix $A_{ij}$ and therefore the local cosmic web 
are entirely determined by $s^{(s)}_i (i=1,2,3)$.
Once the vorticity is generated after shell-crossing, it will then couple to symmetric
contribution, via $s^{(a)}_i$, and affect morphological classification of the local flow.
Therefore, the generation of rotational flow is much more complicated, not merely 
related to vorticity.  Moreover, since $\omega_{ij}$ contributes to both $s_2$ and $s_3$, 
it is possible to generate laminar flow with non-zero vorticity 
(\ref{app_sec:vorticity_diff}).

\subsection{Classification of the Local Flow}
As they are the coefficients of the characteristic equation, the most direct application of 
$s_i$ is the classification of the solutions $\lambda_i$.  
The trajectory around a given point $\vx_0$ can be approximated with
a Taylor expansion, 
\begin{eqnarray}
 \frac{d x_i}{d \tau}(\vx) = u_i(\vx) = u_i(\vx_0) + A_{ij}(\vx_0) x_j + \cdots 
\end{eqnarray}
In the dark-matter comoving frame where $u_i(\vx_0)=0$, ignoring
higher-order terms, one obtains the local trajectory of a fluid 
element via the differential equation
\begin{eqnarray}
\label{eqn:traj_eqn}
 \frac{d x_i}{d\tau} = A_{ij} x_j . 
\end{eqnarray}
Therefore to first order, $\vA$ characterizes the local trajectory of the
fluid element. As will be shown later, neglecting the rotational degree of 
freedom of the coordinate system, various flow morphologies presented by solving
Eq. (\ref{eqn:traj_eqn}) in the canonical form of $\vA$ can be mapped
into invariants space. Therefore, the three-dimensional real parameter space
is then divided into various adjacent regions. 
Among them, the most important criteria is the surface dividing 
the real and complex roots of $\lambda_i$, which is defined by the equation
\begin{eqnarray}
\label{eqn:solution_S1}
 27 s_3^2 + ( 4 s_1^3 - 18 s_1 s_2) s_3 + (4 s_2 ^3 - s_1^2 s_2^2) = 0
\end{eqnarray}
Using the discriminant $s_2 \le s_1^2/3 $ of the above quadratic 
equation, one is able to define real solutions of $s_3$, which we will denote 
as $s_{3a}$ and $s_{3b}$, assuming $s_{3a}\le s_{3b}$.  For real eigenvalues, 
the inequality $s_{3a}\le s_3 \le s_{3b}$ should hold, or there will be complex
eigenvalues otherwise.

In the following of this section, we will review the work of \cite{CPC90}, 
and show how different flow morphologies could be classified according to the 
eigenvalues of $A_{ij}$ and its image in the invariants space. 
Here, we only discuss the most relevant categories in a cosmological context. 
A more complete classification including all degenerating special cases could be
found in the original paper. For convenience, in the Appendix A., 
we also give a list of categories have been discussed, correcting some typos in 
\cite{CPC90}.

As will be seen later, the discussion of flow morphology is possible only in an 
appropriate coordinate system. For potential flow, $A_{ij}$ is symmetric, and therefore 
could always be diagonalized over the field of real numbers. 
\begin{eqnarray}
\label{eqn:Aij_canonical_sym}
\vA =  \vR^{-1} \left( \begin{array}{ccc}
\lambda_1  & \quad  & \quad   \\
\quad & \lambda_2 & \quad  \\
 \quad & \quad & \lambda_3    \\
\end{array}   \right ) \vR, 
\qquad \lambda_i \in \mathbb{R}
\end{eqnarray}
where $\vR$ is a rotation matrix and we assume $\lambda_1 \le \lambda_2 \le \lambda_3$. 
From Eq.\ (\ref{eqn:Aij_canonical_sym}) and the 
definition of $s_i$, the invariants can simply be expressed as 
\begin{eqnarray}
\label{eqn:si_eigval}
s_1 = - \sum_i \lambda_i,  \quad
s_2 = \sum_{i \ne j} \lambda_i \lambda_j,  \quad s_3 = - \prod_i \lambda_i.
\end{eqnarray}
For a general anti-symmetric matrix $\vA$, although in many cases Eq.\ 
(\ref{eqn:Aij_canonical_sym}) and (\ref{eqn:si_eigval})
are still valid over complex field $\lambda_i \in \mathbb{C}$, it is not easy to extract 
geometric information out. Therefore, it is more convenient to rotate the matrix 
into a well-suited frame as described in Sec 2.2.2. 
In some situations, $\vA$ is not diagonalizable. However, 
these are very special cases and are unlikely to be important in cosmological context.

\begin{figure*}[htp]
\begin{center}
\includegraphics[width=0.99\textwidth]{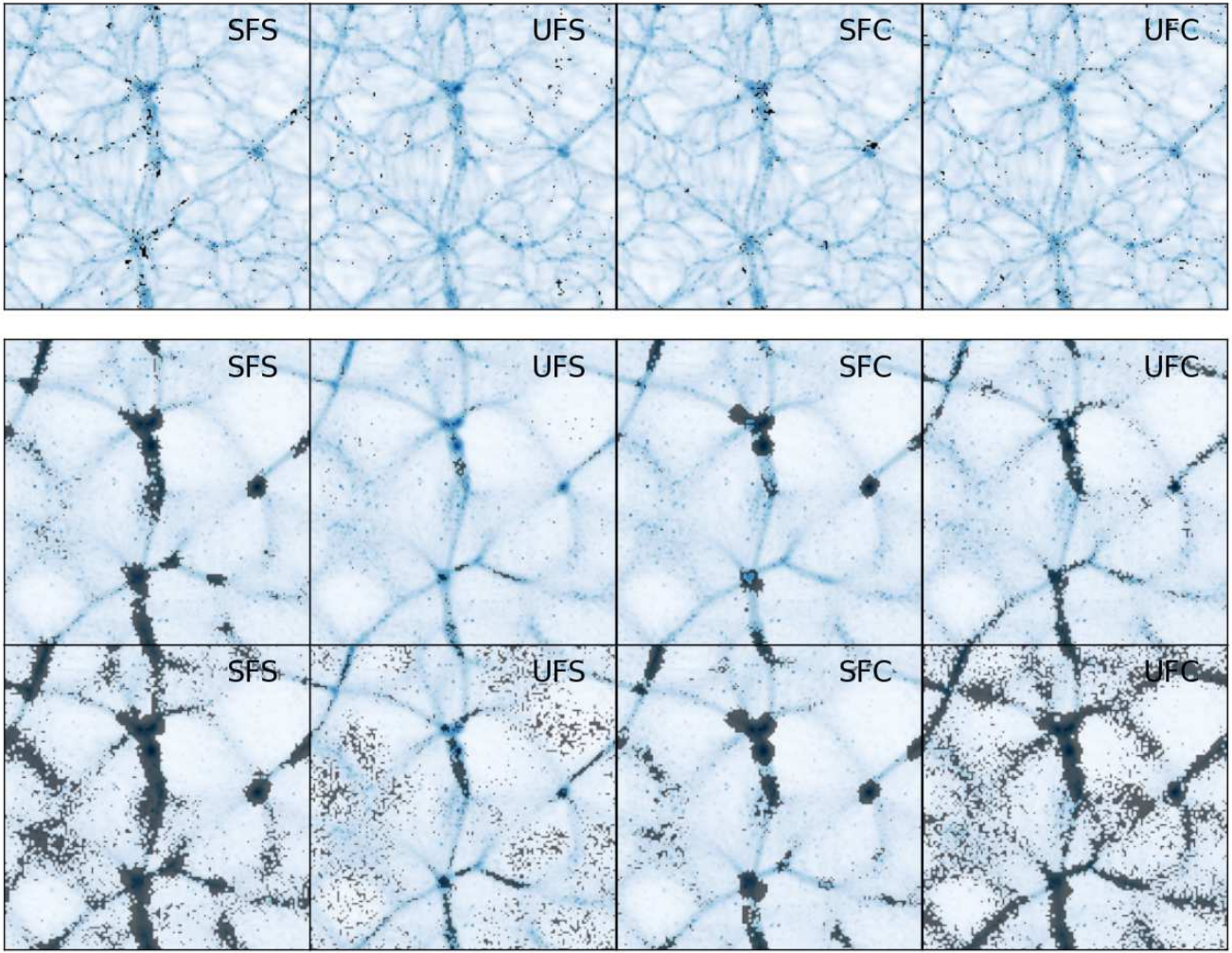}
\end{center}
\caption{ \label{fig:classification_vortical} Four different categories of vortical 
flow in the MIP ensemble, a suite of simulations with the same 
large-scale structure but independent small-scale realizations. 
Every column presents the same morphologies labeled in the upper-right corner of 
each panel.
The top row makes use of only one realization, and therefore show very small fractions
of the region are classified as rotational categories, which is reasonable given 
the resolution of our simulation.
To display more clearly the spatial distribution of various rotational categories, 
we utilize the whole ensemble simulation in the middle and bottom rows, where
much larger area highlights the region where more than $n_{ct}$ out of $64$ realizations
are labeled as a particular rotational category. We adopt $n_{ct}=5$ for the middle
row and $2$ for the bottom. 
The density field of these two are also stacked over all realizations, 
and that is the reason why small-scale structures are smeared out.
}
\end{figure*}

\subsubsection{Potential Flow}
For potential flow, our method categorizes different morphologies according to
the sign of eigenvalues of $A_{ij}$ or corresponding regions in the invariants space. 
In this regime, many similar algorithms have already been proposed to use
either the Hessian matrix of gravitational potential $\nabla_i \nabla_j \Phi$
\citep{HP07a,FR09} or similar to our method, the symmetric part of $A_{ij}$
\citep[`V-web',][]{HM12}. 
Instead of introducing a free parameter of the threshold eigenvalues
$\lambda_{th}$, which is usually adjusted to match the visual impression \citep{HM12}, our 
method simply sets $\lambda_{th}=0$.
However, it is not difficult to incorporate this parameter explicitly. 
One just defines $\lambda^{\pri}=\lambda-\lambda_{th}$ and rewrites the 
characteristic equation (\ref{eqn:chareq}) as a function of $\lambda^{\pri}$, 
\begin{eqnarray}
\label{eqn:chareq_th}
(\lambda^{\pri} + \lambda_{th})^3 + s_1 (\lambda^{\pri} + \lambda_{th})^2 + 
s_2 (\lambda^{\pri} + \lambda_{th}) + s_3 = 0, 
\end{eqnarray}
then discusses the sign of variable $\lambda^{\pri}$ instead. Expanding Eq.\  
(\ref{eqn:chareq_th}), this is equivalent to defining a new set of invariants $s^{th}$, 
\begin{eqnarray}
\label{eqn:s_redfine_th}
s^{th}_1 &=& s_1 + 3 \lambda_{th}\nonumber\\
s^{th}_2 &=& s_2 + 2 \lambda_{th} s_1 + 3 \lambda_{th}^2 \nonumber\\
s^{th}_3 &=& s_3 + \lambda_{th} s_2 + \lambda_{th}^2 s_1 + \lambda_{th}^3 .
\end{eqnarray}
Then all classification conditions of the invariants that will be introduced in the 
following remain the same.

Since the symmetric matrix $A_{ij}$ can be diagonalized into canonical form 
(\ref{eqn:Aij_canonical_sym}), there will exist three eigen-planes, that 
contain solution trajectories.
These trajectories may look like nodes and saddles on the plane 
\footnote{In this paper, we adopt the same terminology describing the kinematic 
morphologies as in turbulence literature, e.g. \cite{CPC90}, where the
nodes or saddles denote the configuration of physical flow trajectory in the eigen-plane.
These do not necessarily correspond to `nodes' that correspond to halos or clusters, in a density field classification. }, 
depending on the sign of the eigenvalues.
Particularly when all eigenvalues $\lambda_i<0$, then three planes contain nodes.
So the mass elements would all flow toward the position of interest $\vx_0$ in each eigen-planes,
which is more likely to occur in the over-dense region, around halos.
The mapping from $\lambda_i<0$ in eigenvalue space into invariants space is 
straightforward, since from Eq. (\ref{eqn:si_eigval}), we see $s_i>0$ for 
all $i=(1,2,3)$.
Combined with the real solution condition (Eq.\ \ref{eqn:real_cond}), this type 
corresponds to a region in invariants space, 
\begin{eqnarray}
 s_1 >0, ~ 0<s_2<s_1^2/3, ~ \max(s_{3a},0)<s_3<s_{3b}. ~
\end{eqnarray}
As illustrated in Fig.\ (\ref{fig:flow_category}), the streamlines in this region are 
all `stable nodes' in all three eigen-planes, and therefore it will be denoted as 
`$\mathcal{SN/SN/SN}$' in the following.

On the other hand, when all eigenvalues $\lambda_i > 0$, one still gets three nodes,
but the flow will all head outward from the origin, which most likely would happen in under-dense region, in the void.
Similarly from Eq. (\ref{eqn:si_eigval}), $s_1, s_3<0$ and $s_2>0$, and with 
real root condition, it corresponds to region in invariants space
\begin{eqnarray}
s_1<0, ~  0<s_2<s_1^2/3, ~ s_{3a}<s_3<\min(s_{3b}, 0). ~
\end{eqnarray}
And we will denote this type as `$\mathcal{UN/UN/UN}$' where `$\mathcal{UN}$' 
stands for `unstable node'.

If the matrix $A_{ij}$ becomes indefinite, i.e. with both positive and negative eigenvalues, 
then it possess saddle points in two eigenvector planes. As shown in Fig. 
(\ref{fig:flow_category}), in the eigen-plane with saddle point, the flow will 
approach inward from one direction and depart toward the other without 
passing through the origin.
Particularly when the smallest two eigenvalues $\lambda_1, \lambda_2$ are negative, 
while  $\lambda_3$ positive, two saddle points reside in the planes spanned by 
$\vv_1 - \vv_3$ and $\vv_2-\vv_3$, where $\vv_i$ is the eigenvector corresponding to
$\lambda_i$, and the node in plane $\vv_1 - \vv_2$ is stable, i.e. mass elements flow
inward. 
Therefore, we will name this category as `$\mathcal{ SN/S/S}$', where the last two
`$\mathcal{S}$' denote `saddle point'. 
From the illustration of the trajectory in Fig. (\ref{fig:flow_category}), 
this type resembles the situation when the mass element flow along the 
filamentary structures.
In invariants space, the sum of $\lambda_i$ could be either positive or 
negative depending on the relation between $|\lambda_1+\lambda_2|$ and $\lambda_3$. 
If $s_1\ge 0$, i.e. $|\lambda_1+\lambda_2| \ge \lambda_3$, the flow changes
faster in the node plane $\vv_1-\vv_2$, and causes the net compression of the fluid element. 
While if $s_1< 0$, it happens the opposite way. In summary, we have
\begin{eqnarray}
\qquad && ~~ s_1\ge 0, \qquad s_2<s_1^2/4, \qquad s_{3a}<s_3<0, \nonumber\\
or && ~~ s_1 < 0, \qquad s_2<0, \qquad s_{3a}<s_3<0.
\end{eqnarray}
When only the smallest eigenvalue $\lambda_1$ is negative, and $\lambda_2, 
\lambda_3$ positive, then two saddle points reside in the planes spanned by 
$\vv_1 - \vv_2$ and $\vv_1-\vv_3$, and the node in the plane $\vv_2 - \vv_3$ is
unstable since $\lambda_{2,3}>0$. This corresponds to the case when matter flow towards 
a wall structure. Again in the invariants space, we'll also have two conditions
\begin{eqnarray}
\qquad && ~~ s_1\le 0, \qquad s_2<s_1^2/4, \qquad 0<s_3<s_{3b}, \nonumber \\
or && ~~ s_1 > 0, \qquad s_2<0, \qquad 0<s_3<s_{3b}.
\end{eqnarray}
Similarly, $s_1\le 0$ corresponds to faster velocity change in the node 
plane $\vv_2-\vv_3$, and a net expansion of the fluid element. If $s_1 >0$, 
the fluid element contracts.

\begin{figure}[htp]
\begin{center}
\includegraphics[width=0.48\textwidth]{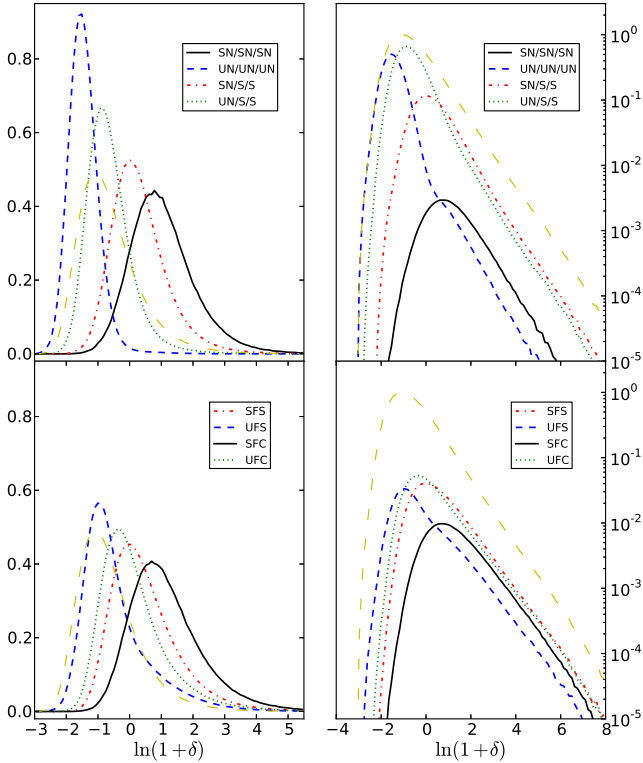}
\end{center}
\caption{ \label{fig:density_dist} 
{\it Left:} the PDF of density $1+\delta$ for various flow morphologies, drawn 
from the `$Sing100$' simulation.
{\it Right:} same as left panels but all curves have been rescaled by the volume 
fraction of each morphology type, so that the maxima of entire density field 
(long-dashed line)  equal unity.
{\it Upper:} distribution of various potential flow. 
{\it Lower:} distribution of rotational flow. 
}
\end{figure}

\subsubsection{Vortical Flow}
After shell crossing, the anti-symmetric part $\omega_{ij}$ of the matrix 
$A_{ij}$ will be generated. In terms of real numbers, the canonical form of 
$A_{ij}$ can be expressed as
\begin{eqnarray}
\label{eqn:Aij_canonical_asym}
\vA =  \vR^{-1} \left( \begin{array}{ccc}
a  & -b & \quad   \\
b  & a & \quad  \\
 \quad & \quad & c    \\
\end{array}   \right ) \vR, 
\end{eqnarray}
where $a, b, c$ are all real, and the complex eigenvalue 
$\lambda_{1,2} = a \pm ib$, $\lambda_3 = c$. 
Then the invariants simply read as 
\begin{eqnarray}
\label{eqn:vort_si_canonical}
s_1 = -(2 a + c),  \quad s_2 = a^2 + b^2 + 2 a c, \nonumber \\
s_3 = - c(a^2 + b^2). \qquad \qquad \quad
\end{eqnarray}
In this case, there will exist one plane, corresponding to eigenvalues 
$\lambda_{1,2}$, that contains the solution trajectory, which can be 
expressed as
\begin{eqnarray}
  r = r_0 e^{m \alpha }
\end{eqnarray}
in polar coordinates $(r, \alpha)$. Here the factor $m=a/b$ denotes the rate 
of spiraling, and $r_0$ is a constant depending on the initial condition. 
Along the direction of $\lambda_3$, the behavior of the trajectory is then
controlled by the value of $c$.
 
Depending on the value of $m$ and $c$, and neglecting degenerate cases, 
there remain four major categories of rotational flow.
For negative $m$, the trajectory spirals inwards to the origin of the 
plane, which will be called `stable focal flow' in the following. 
If $c>0$, mass elements flow away from the plane, a situation denoted `$\mathcal{SFS}$', i.e. `stable focal stretching.'
If $c<0$, mass elements flow towards the plane, and we will call it as
`stable focal compressing', abbreviated as `$\mathcal{SFC}$'. 
For positive $m$, the trajectory spirals outwards from the origin, and the flow is called 
`unstable focal flow'. 
Depending on the value of $c$, the last two flow types are called 
`unstable focal stretching', abbreviated as `$\mathcal{UFS}$' for $m>0, c>0$; 
`unstable focal compressing', abbreviated as `$\mathcal{UFC}$' for $m>0, c<0$.
The detailed criteria for vortical regions are listed in Appendix A.

\subsection{Cosmic Web and Kinematic Morphologies}
To show the relationship between large-scale-structure morphologies and the 
kinematic classification, we highlight different categories of 
flow patterns of `$Sing100$' simulation at redshift $z=0$ and compare with the 
density field in Fig. (\ref{fig:classification_potential}). 
Since the process of cosmic web formation is largely associated with potential
flow, and to avoid disturbance from rotational degree of freedom, we smooth the 
velocity field with a Gaussian filter with smoothing length varying from 
$R=0.2\Mpc/h$ to $1.5\Mpc/h$. 
Notice that different categories are labeled by their kinematic names such as
`$\mathcal{SN/SN/SN}$', instead of the more familiar `halo', since by setting 
$\lambda_{th}=0$, our morphology classification does not necessarily reproduce
the best visual structures compared with `V-web' \citep{HM12}. 
Since our method includes both potential and rotational morphologies, cosmic-web 
structures, especially filaments, defined with irrotational flow could only be identified 
with relative large smoothing length. 
However, with an appropriate smoothing, e.g. $R=0.8\Mpc/h$ as shown in the lower 
left panel, one does see a good visual correspondence between filamentary cosmic 
web in the density
(first panel) and the corresponding kinematic categories `$\mathcal{SN/SN/SN}$' (halo-like) 
and `$\mathcal{SN/S/S}$' (filament-like).
This mainly reflects the connection between the velocity potential $\psi$ and 
gravitational potential $\Phi$ on large scales. 
As one performs the smoothing more aggressively, features at small scale vanish as expected.
However, unlike the wall/sheet structure identified in \cite{HM12}, our 
wall-like `$\mathcal{UN/S/S}$' type fills almost half of the simulation volume. 
Although more studies are desired, we tend to believe this is a combined effect of 
both the larger Eulerian volume filling factor for underdense regions and the 
relative larger volume in the invariants space, as could be seen in Fig. (\ref{fig:flow_category}).
From Fig. (\ref{fig:classification_potential}), we could also see that the large-scale
structure identified kinematically is reliable with various degree of smoothing, 
although the rotational categories, shown in yellow, is reduced dramatically as one 
increases the smoothing length.

To better understand these rotational regions in Fig.\ (\ref{fig:classification_potential}), 
we use the MIP $N$-body ensemble simulations and show the spatial distribution of four 
vortical flow categories in Fig.\ (\ref{fig:classification_vortical}). 
One benefit of using the MIP simulation is that, since vorticity is only generated at
small scales after shell-crossing, in a single realization of a moderate resolution 
simulation, one will not have many pixels classified as rotational. 
Moreover, because the classification criteria is mutually exclusive, each pixel can 
only be categorized as one of them.
On the other hand, with the full suite of MIP simulation, one can explore the ensemble probability of generating one particular morphology given the 
large-scale environment.
For this purpose, we first make a complete classification of all realizations, and 
then highlight the region where more than $n_{ct}$ out of $64$ are categorized  
as a certain type. By doing this, each pixel of the simulation could be multi-valued.
The result is shown in Fig. (\ref{fig:classification_vortical}), where the first row
illustrate the vortical classification of an arbitrary realization while 
the middle and bottom row are combined classification with $n_{ct}=5$ and $2$ respectively. 
The underlying density distribution of the last two rows are produced by stacking all
$64$ realizations, therefore lacking the small-scale structure compared with the first. 
Similar to the irrotational morphologies, the most distinctive feature of the 
figure is that various flow categories display different spatial
distributions. 
For example, `$\mathcal{SFC}$' resides around knots
and `$\mathcal{SFS}$' and `$\mathcal{UFC}$' are located along filaments; this  
is especially clear for lower $n_{ct}$. 
Furthermore, as we will discuss further below, the 
direction of the vorticity of different categories also vary among different
categories.

Besides the spatial distribution, we also plot in Fig. (\ref{fig:density_dist})
the probability density function (PDF) of matter density for various irrotational 
flow categories in the upper panels and vortical types in the lower.
While the PDF is displayed at left, we also present curves rescaled by 
the volume fraction of each morphology type at right.
From left panels, one notices the locations of PDF maxima of each category in 
$(1+\delta)$-axis, $\delta_{max}$, exhibit a sequence consistent with their 
cosmic web morphologies.  
That is, the PDF of type `$\mathcal{SN/SN/SN}$' peaks at higher density value
$\delta$ than that of type `$\mathcal{SN/S/S}$'; we identify these as halo and 
filament respectively; following these are wall/sheet and void types. 
On the other hand, as shown in the lower panel, $\delta_{max}$ of 
vortical types are all greater than that of the entire field (long-dashed line), 
which is reasonable since they are only generated after shell-crossing in denser regions.  
Although the differences of peak positions among all vortical categories are narrower
than for potential types, they are still consistent with the spatial distribution 
from Fig\. (\ref{fig:classification_vortical}): the type `$\mathcal{SFS}$' around 
knots are highest-density, then follow `$\mathcal{SFS}$' and `$\mathcal{UFC}$', which 
trace filaments. 
From the right panels, one could further see the relative volume abundance of each category. 
Since the distributions are volume weighted in Eulerian space, higher density regions are expected to be less abundant. This explains the lower amplitude of category `$\mathcal{SN/SN/SN}$' and `$\mathcal{SFC}$' in the figure. Also, as consistent with the visual 
impression of Fig.\ (\ref{fig:classification_potential}), more regions are classified as 
`$\mathcal{UN/S/S}$' instead of `$\mathcal{UN/UN/UN}$'.

\begin{figure}[htp]
\begin{center}
\includegraphics[width=0.49\textwidth]{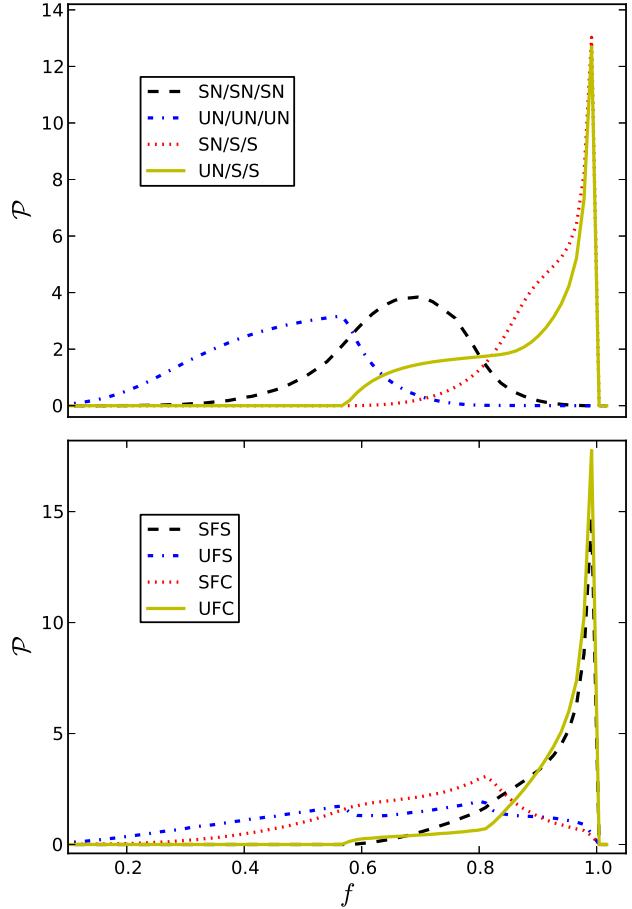}
\end{center}
\caption{ \label{fig:anist_shear} The fractional anisotropy defined in Eq.\
(\ref{eqn:frac_anist}) for various flow morphologies from `$Sing100$' simulation. 
{\it Upper:} Potential flows; {\it lower:} rotational flows. 
}
\end{figure}

To further examine our invariants-based cosmic web classification, we also investigate 
the isotropy of each category, since to some level halos and voids should be more isotropic
than filaments and walls. 
For the same purpose, \cite{L13} measured the fractional anisotropy for their shear based
`V-web' classification, 
\begin{eqnarray}
\label{eqn:frac_anist}
f_a= \frac{1}{\sqrt{3}} \sqrt{\frac{\sum_{i<j}(\lambda_i-\lambda_j)^2}
{\sum_i \lambda_i^2} },
\end{eqnarray}
where $\lambda_i$ are eigenvalues of symmetric tensor $\theta_{ij}$. 
It takes values between zero and unity, with $f_a=0$ for totally isotropic 
expansion/contraction and $f_a=1$ for anisotropic motion. 
Due to their selected eigenvalue threshold $\lambda_{th}=0.44$, \cite{L13} 
find that voids in their classification have the highest anisotropy, while filaments
and walls show a broad distribution of $f_a$. 
However, as shown in the upper panel of Fig.\ (\ref{fig:anist_shear}), in our 
classification, both filaments and walls exhibit highest anisotropy, whereas halo
and void are relatively isotropic. Especially, contrary to \cite{L13}, the voids here 
show the lowest anisotropy.
This suggests that, although an fine-tuned $\lambda_{th}$ might produce a better web 
structure visually, some of its results may be harder to interpret physically.

In the lower panel of Fig.\ (\ref{fig:anist_shear}), we also display the same quantities
for vortical flows. Clearly, one sees that the filament-tracing types `$\mathcal{SFS}$' and
`$\mathcal{UFC}$' are highly anisotropic. This suggests that even after generating rotational 
degree of freedom, these fluid elements in filamentary environment mostly still follow the
 similar shearing movement. On the contrary, other categories show relatively flat distributions.

\section{Gravitational Evolution Before Shell-crossing}
In the next two Sections, we will theoretically investigate the evolution of the velocity gradient tensor both before and after shell-crossing. 
Before shell-crossing, one is able to neglect the stress tensor in the Euler 
equation and consider the so-called dust model \citep{P80,BCGS02},  
\begin{eqnarray}
\label{eqn:euler_eq}
\frac{\partial \vu}{\partial \tau} + \mH(\tau) \vu(\vx, \tau) + 
\vu(\vx, \tau)\cdot \nabla \vu (\vx, \tau) =  -\nabla \Phi(\vx,\tau) \qquad
\end{eqnarray}
where $\mH=d \ln a /d\tau$, 
and $\Phi$ is the Newtonian potential, which satisfies the Poisson equation
\begin{eqnarray}
\label{eqn:Poisson_eq}
\nabla^2 \Phi(\vx, \tau) =  \frac{3}{2} \Omega_m(\tau) \mH^2(\tau) \delta(\vx, \tau).
\end{eqnarray}
And one can then close the system with the matter continuity equation,
\begin{eqnarray}
\label{eqn:continuity_eq}
\frac{\partial \delta}{\partial \tau} + \nabla \cdot [(1+\delta)\vu] = 0.
\end{eqnarray}
Eq. (\ref{eqn:euler_eq})-(\ref{eqn:continuity_eq}) are basic equations of 
large-scale structure in Newtonian cosmology.  For potential flow, it is sufficient 
to take the velocity potential $\phi(\vx)$, or equivalently the divergence 
$-s_1(\vx) = \nabla^2 \phi(\vx)$, as the only dynamical degree of freedom of $u_i(\vx)$, 
since other quantities including $u_i(\vx)$ and $A_{ij}(\vx)$ could be recovered by 
explicit spatial derivatives. 
For our purpose, however, the flow morphology classification at given position $\vx$ 
requires more information of tensor $A_{ij}(\vx)$ than a single scalar. 
Moreover, instead of examining the tensorial field $A_{ij}(\vx)$, in the following,
we would like to follow a fluid element, and investigate its Lagrangian morphological
evolution.
To proceed, one takes the gradient of Euler equation (\ref{eqn:euler_eq}) and obtain the 
Lagrangian evolution equation of the tensor $A_{ij}$,
\begin{eqnarray}
\label{eqn:Aij_evl_full}
\frac{d A_{ij}}{d \tau} + \mH(\tau) A_{ij} + A_{ik} A_{kj} = \varpi_{ij},
\end{eqnarray}
where we have written the Lagrangian total derivative 
$d/d\tau = \partial/\partial \tau + \vu\cdot \nabla $, and defined
$\varpi_{ij}(\vx, \tau) = -\nabla_i \nabla_j \Phi(\vx, \tau)$.
Here, the Lagrangian frame enables us to trace the change of flow morphology of 
a mass element, and as will be shown later, in some cases it will simplify the 
dynamical equations.
Eq.\ (\ref{eqn:Aij_evl_full}) holds for each element of the gradient tensor. 
To marginalize over these degrees of freedom, one is interested in the evolution 
equation of the invariants $s_i$. 
Multiplying both sides of Eq.\ (\ref{eqn:Aij_evl_full}) by $\vI, \vA$ and $\vA^2$ 
respectively, and then taking the trace, one obtains the dynamical equation of 
$s_1, s_2$, and  $s_3$,
\begin{eqnarray}
\label{eqn:inv_dyn_full}
&& \frac{d }{d\tau} s_1 + \mH(\tau)s_1 -  s_1^2 +  2 s_2  = -\varpi 
 \nonumber \\
&& \frac{d}{d\tau} s_2  + 2\mH (\tau)s_2  - s_1 s_2  + 3 s_3  = - s_1 
\varpi  - \varpi_{A}   \nonumber \\
&& \frac{d}{d\tau} s_3  + 3\mH(\tau) s_3 -s_1 s_3 =  -s_2 \varpi  - s_1 \varpi_{A}
 - \varpi_{A^2}
\end{eqnarray}
with source terms defined as $\varpi=\varpi_{ii}=-\nabla^2 \Phi$, 
$\varpi_A=\varpi_{ij}A_{ji}$ and  $\varpi_{A^2} = \varpi_{ik}A_{kj}A_{ji}$.
Here we have used the following identities to simplify the expression,
\begin{eqnarray}
\tr[\vA^2] &=& A_{ij}A_{ji}  = s_1 ^2 - 2 s_2  \nonumber\\
\tr[\vA^3] &=& A_{ij}A_{jk}A_{ki} = -s_1^3 + 3 s_1 s_2 - 3 s_3 \nonumber \\
\tr[\vA^4] &=& A_{ij}A_{jk}A_{kl}A_{li}  = s_1^4 - 4 s_1^2 s_2 + 4 s_1 s_3 + 2 s_2^2.
\nonumber \\
\end{eqnarray}
The final equation is derived with the help of the Cayley-Hamilton theorem.
Eq.\ (\ref{eqn:inv_dyn_full}) depends on the full tidal tensor $\varpi_{ij}$ and its 
coupling with $A_{ij}$, and hence is non-trivial to solve.  The scalar $\varpi$ is simply proportional to the density, and can be obtained via the continuity equation,
$\varpi= - (3/2) \Omega_m \mH \delta $.
To close Eq.\ (\ref{eqn:inv_dyn_full}), one has to supplement the evolution of tidal 
tensor, either new equation(s) or known solution from other technique.
On the other hand, as will be seen shortly, Eq.\ (\ref{eqn:inv_dyn_full}) 
is dramatically simplified in the Zel'dovich approximation.

\begin{figure}[htp]
\begin{center}
\includegraphics[width=0.49\textwidth]{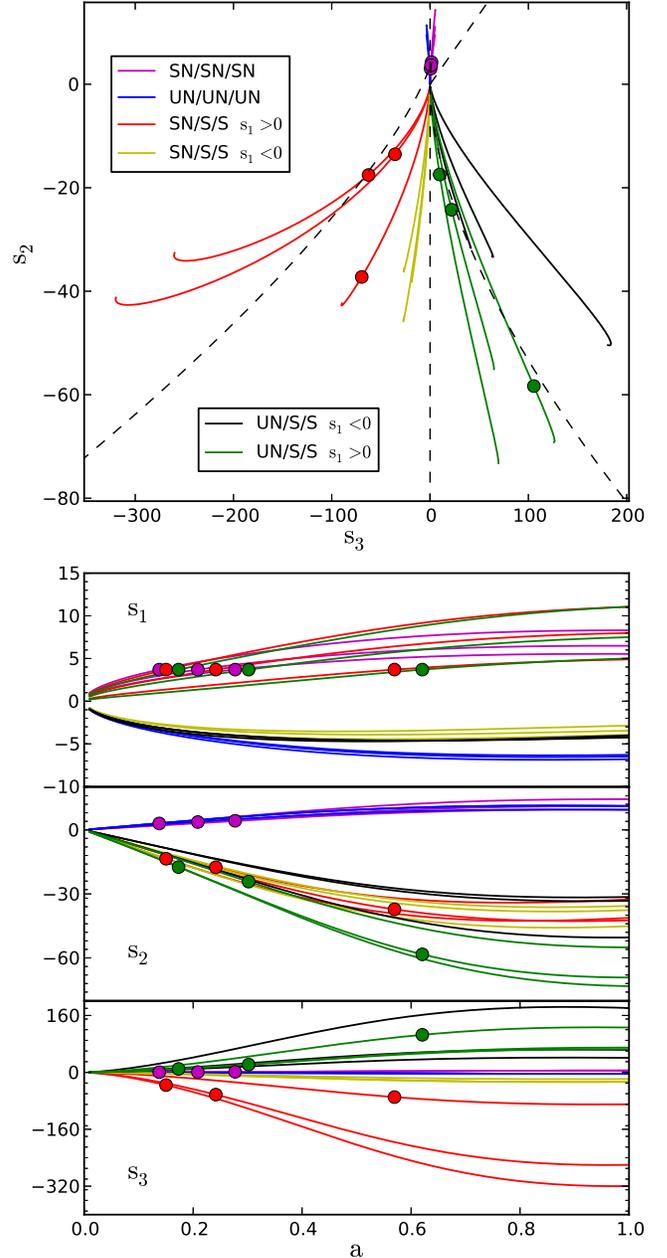}
\end{center}
\caption{ \label{fig:inv_za_evolution} Dynamical evolution of the invariants in the Zel'dovich 
approximation, for various initial conditions. Different colors represent various initial
morphologies. 
{\it Upper: } Evolution trajectory in the two-dimensional invariants space $s_3-s_2$. Black
dashed lines separate different categories at fixed $s_1$, which is also marked by solid
circles on each trajectories.
{\it Lower: } Time evolution of each invariant. 
}
\end{figure}

\subsection{Dynamical evolution of invariants in the Zel'dovich Approximation}
In Lagrangian dynamics, the mass element moves in the gravitational potential
along the trajectory \citep{ZA70,BCGS02}
\begin{eqnarray}
\vx(\vq, \tau)=\vq+\vpsi(\vq, \tau),
\end{eqnarray}
from initial Lagrangian position $\vq$. 
To first order, i.e.\ the Zel'dovich approximation (ZA) \citep{ZA70}, the displacement field
 $\Psi(\vq, \tau)$ is given by
\begin{eqnarray}
\nabla \cdot \Psi(\vq, \tau) = - D(\tau) \delta(\vq),
\end{eqnarray}
where $D(\tau)$ the linear growth factor of density perturbation.
In Eulerian space, this is equivalent to replacing the Poisson equation with 
\citep{MS94,HB96,BCGS02}
\begin{eqnarray}
\label{eqn:za_assump}
u_i(\vx, \tau) &=& - \frac{2 f(\tau)}{3 \Om(\tau) \mH(\tau) } \nabla_i \Phi(\vx, \tau),
\end{eqnarray}
which then closes the system together with Euler equation (\ref{eqn:euler_eq}). 
Here, $\Omega_m(\tau)$ is the matter density fraction at epoch $\tau$, and 
$f=d\ln D/ d\ln a$ is the growth rate.  
In this approximation, one finds that $\varpi, \varpi_A$ and $\varpi_{A^2}$ equal 
\begin{eqnarray}
\label{eqn:varphi_za}
\varpi &=& - \mH(\tau) \mL(\tau) s_1 \nonumber \\
\varpi_A &=& \mH(\tau) \mL(\tau) (s_1^2 - 2 s_2 ), \nonumber \\
\varpi_{A^2} &=& \mH(\tau) \mL(\tau) (-s_1^3 + 3 s_1 s_2 - 3 s_3 ).
\end{eqnarray}
where we have defined $\mL(\tau) = 3 \Om(\tau)/ 2f(\tau)$.

Substituting  Eq. (\ref{eqn:varphi_za}) back into Eq. (\ref{eqn:inv_dyn_full}), one obtains
the full dynamical equations of the invariants $s_i$ in the Zel'dovich approximation. 
They can be further simplified by defining the rescaled velocity 
$\bvu=\vu/D^{(v)}$ where $D^{(v)}(\tau)=dD/d\tau=\mH f D$, and change the time 
variable $\tau$ in Eq. (\ref{eqn:euler_eq}) into the linear growth rate $D$.  Then, 
Euler equation reads \citep{SZ89}
\begin{eqnarray}
\bvu^{\pri} =  \frac{d \bvu}{d D}= 
\left( \frac{\partial}{\partial D} + \bvu\cdot \nabla \right ) \bvu= 0,
\end{eqnarray}
where the prime denotes the Lagrangian derivative $d/dD$, and we have also used
differential equation of the linear growth rate \cite{P80,BCGS02}
\begin{eqnarray}
\frac{d^2 D(\tau)}{d^2\tau} +\mH(\tau) \frac{d D(\tau)}{d\tau} = \frac{3}{2} 
\Omega_m(\tau) \mH^2(\tau) D(\tau).
\end{eqnarray}
For the velocity gradient tensor, we similarly define the rescaled quantity
$\bA_{ij}=A_{ij}/D^{(v)}$, and
\begin{eqnarray}
\frac{d \bA_{ij} }{d D} + \bA_{ik}\bA_{kj} = 0.
\end{eqnarray}
From the definition of various categories in Appendix A, all boundaries classifying
different flow categories are invariant if the velocity is scaled with any positive number. 
Therefore from the definition in Eq.\ (\ref{eqn:inv_def}), we define the rescaled invariants 
$\bs_1, \bs_2, \bs_3$ as
\begin{eqnarray}
\bs_i(\tau) = \frac{s_i(\tau)}{[D^{(v)}]^i}, \qquad i \in \{1,2,3\}. \quad
\end{eqnarray}

\begin{figure*}[htp]
\begin{center}
\includegraphics[width=1.\textwidth]{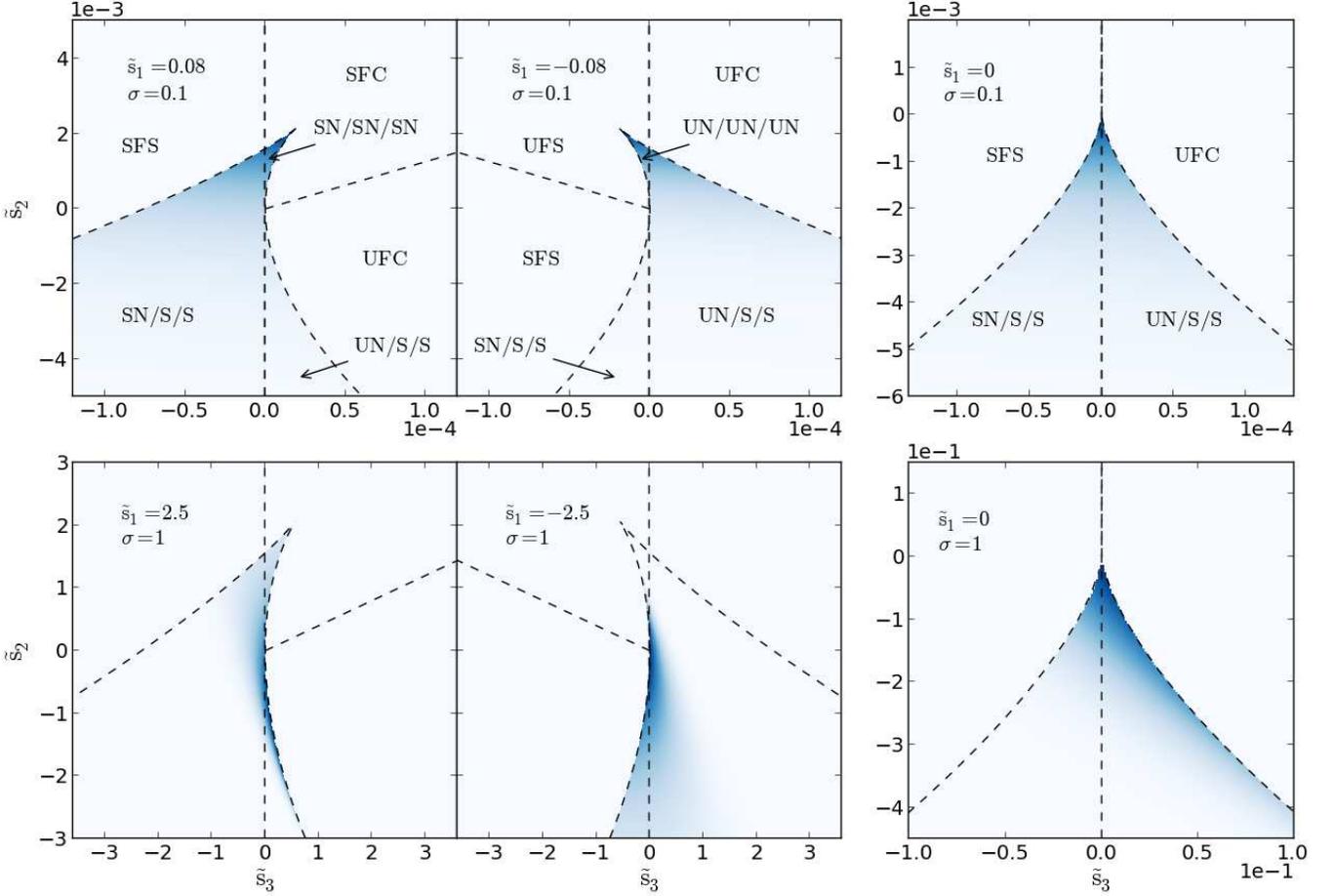}
\end{center}
\caption{ \label{fig:PDF_za_theory} Theoretical probability density of $\ts_3-\ts_2$
in the Zel'dovich approximation at $\sigma=0.1$ and $1$, for various $\ts_1$ 
($\pm0.08$, $0$ for $\sigma=0.1$, and $\pm 2.5$, $0$ for $\sigma=1$),
assuming $\sigma=1$. 
The dashed lines represent boundaries among classifications, and the color 
scheme is the same within each plot. Notice that axes scales differently among
different figures. 
}
\end{figure*}

One can then derive a set of solvable ordinary differential equations of  
reduced invariants:
\begin{eqnarray}
\label{eqn:inv_za_evolution}
\bs_1^{\pri} - \bs_1^2 + 2\bs_2 &=& 0, \nonumber \\
\bs_2^{\pri} - \bs_1 \bs_2 + 3\bs_3 & =& 0, \nonumber \\
\bs_3^{\pri} - \bs_1 \bs_3 & = &0.
\end{eqnarray}

In Fig. (\ref{fig:inv_za_evolution}), we plot the numerical solution of the evolution 
of invariants for various sets of initial conditions, where different colors indicate 
distinct initial flow morphologies.
The upper panel illustrates the time evolution in the projected $s_3-s_2$ plane, starting 
from the vicinity around zero point, since $s_i$ roughly grows as $(D^{(v)})^i$ to the
lowest order.
The thin dashed boundaries are drawn at a certain $s_1$, which are marked as a solid circle 
on top of each trajectory. 
Hence even though some of the trajectories seem to visually pass through these boundary
lines, it doesn't mean the morphologies have really changed, simply because the 
boundaries also evolve with time.  
In the lower part, we also plot the time evolution of each invariant with
the same color scheme.

\subsection{Initial Condition and Evolution of Probability Distribution}
The fact that we obtain a set of ordinary differential equations 
(Eq. \ref{eqn:inv_za_evolution}) in the Zel'dovich approximation reflects its 
local assumption of the dynamical evolution, which means that the flow morphologies, 
and other properties of mass elements as well, decouple from the nearby
environment, and are entirely determined by the initial conditions. 
In the Zel'dovich approximation, it is straightforward but still physically relevant to study the evolution of the distribution of the invariants $s_i$ away from the initial conditions.

We first notice that the velocity gradient tensor 
is closely related to the tensor $\xi_{ij} = \partial \Psi_i /\partial q_j (\tau)$, 
\begin{eqnarray}
\label{eqn:Aij_za}
\tA_{ij}&=&  \frac{A_{ij} (\tau) }{d\ln D/d\tau (\tau)}= \frac{\partial  \Psi_i(\vq, \tau) }
{\partial x_j} = \frac{\partial \Psi_i(\vq, \tau)}{\partial q_k}  (J^{-1})_{kj}. \nonumber\\
\end{eqnarray}
For convenience, here we define another reduced quantity $\tA$, where $d\ln D/d\tau =
 D^{(v)}/D= \mH f$, and $J_{ij}$ is the Jacobian matrix 
from Lagrangian $\vq$ to Eulerian space $\vx$, where  $J_{ij} = \partial x_i/ q_j = \
\delta_{ij} + \xi_{ij}(\tau). $
We also assume that the Jacobian is invertible, which is true before shell-crossing. 
Concentrating on the initial conditions, components of $\partial \Psi_i(\vq,\tau)/
\partial q_j$ are small compared to unity, so $J_{ij}\sim I_{ij}$ and 
$\tA_{ij} \sim \xi_{ij}$. In the following, we will write all relevant quantities
in this limit with superscript $^{(\xi)}$, such as $\tA^{(\xi)}_{ij}$ and 
$\ts^{(\xi)}_i$. 
Writing the diagonal representation of $\tA^{(\xi)}_{ij}$ as
\begin{eqnarray}
\tA^{(\xi)}_{ij} = diag (\lambda_1, \lambda_2, \lambda_3 ),
\end{eqnarray}
with the help of the distribution function of ordered eigenvalues $\mP(\{\lambda_i\})$ 
for a Gaussian field \citep{D70}, one could derive the distribution function
of invariants $\ts^{(\xi)}_i$ \citep{BBKS86,PG09,PG10}
\begin{eqnarray}
\mP \left (\left \{ \ts^{(\xi)}_i \right\} \right) &=&  \frac{15^3}{8\pi \sqrt{5} 
~ \sigma^6 } e^{ -\frac{3}{2\sigma^2} \left[ 2\left(\ts^{(\xi)}_1 \right)^2 - 5 
\ts^{(\xi)}_2 \right]} \nonumber \\
& & \times \ \ Real\left(\left\{ \ts^{(\xi)}_i \right\} \right).
\end{eqnarray} 
Here, the variance $\sigma$ of the density fluctuation is
\begin{eqnarray}
\sigma^2 = \frac{1}{2\pi^2}\int dk ~k^2 P(k) W^2(kR),
\end{eqnarray}
and $W(kR)$ is a window function with smoothing length $R$. 
The function $Real(\{\ts_i\})=\Theta(\ts_1^2-3\ts_2 ) Rec(\ts_3, \ts_3^a, \ts_3^b)$ 
defines the region in $\ts_i$ space with real eigenvalue solutions, 
where $\Theta(x)$ is Heaviside step function, and $Rec(x,x_1,x_2)$ is a rectangular
function that equals one if $x_1<x<x_2$, and zero otherwise.

As tensor $\xi_{ij}(\tau)$ grows large enough that the effect of the Eulerian gradient becomes 
non-negligible, the matrix $\tA_{ij}$ becomes non-Gaussian even in the Zel'dovich
approximation. 
With the definition Eq.\ (\ref{eqn:Aij_za}), we again consider the diagonalized matrix
\begin{eqnarray}
\label{eqn:tA_eig}
\tA_{ij} = diag (\teta_1, \teta_2, \teta_3 ).
\end{eqnarray}
In general, the eigenvector system defined by $\tA_{ij}$ and $\xi_{ij}$ will differ.  However, since the addition and the inverse of a non-singular matrix does not change the eigenvectors, 
we have
\begin{eqnarray}
 \teta_i = \frac{\lambda_i}{1+\lambda_i}, \qquad 1+\lambda_i \ne 0.
\end{eqnarray}

This allows us to estimate a Lagrangian-space (i.e. mass-weighted) 
PDF of the invariants of $\tA_{ij}$
\begin{eqnarray}
\mP \left (\left\{ \ts_i \right\} \right) &=& \mP \left (\left\{ \lambda_i \right\} \right)
\left | \frac{\partial \lambda_i}{ \partial \teta_k}
\frac{\partial \teta_k}{ \partial \ts_j} \right| 
= \frac{15^3}{8\pi\sqrt{5 }~\sigma^6 ~T^4}  Real( \{\ts_i \}) \nonumber\\
&& \times \exp \biggl[ \frac{-3(\ts_1+ 2\ts_2+3\ts_3)^2 }{\sigma^2 T^2} 
 + \frac{15(3\ts_3+\ts_2)}{2\sigma^2 T } \biggr],  \nonumber\\
\end{eqnarray}
where we have defined the notation
\begin{eqnarray}
T =  1+ \sum_i \ts_i =1+ \ts_1 + \ts_2+ \ts_3.
\end{eqnarray}
$Real(\ts_i)$ is the same function defined previously.

\begin{figure*}[htp]
\begin{center}
\includegraphics[width=0.92\textwidth]{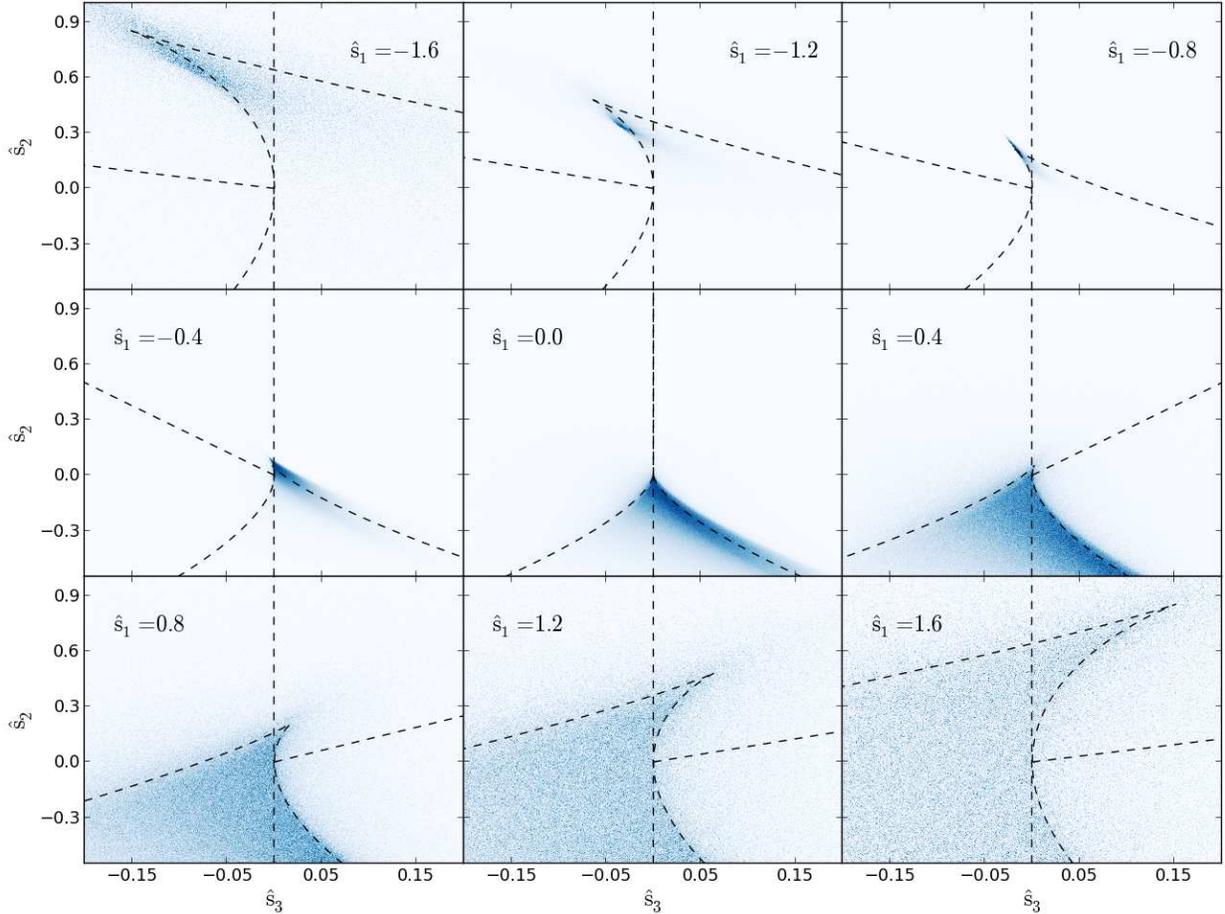}
\end{center}
\caption{ \label{fig:pdf_2d_simulation} Two-dimensional probability of $\hat{s}_3$ and
$\hat{s}_2$ for nine different bins of $\hat{s}_1$, 
measured from simulations at $z=0$. Note that the hat quantities $\hat{s}_i$
denote the rescaling of the tensor $A_{ij}$ Eq. (\ref{eqn:Aij_rescale}) so that invariants
are confined within given rectangular region. From left to right, 
$\hat{s}_1$ goes from $-1.6, -1.2, -0.8$ in the first row, to $-0.4, 0, 0.4$ in the middle, and then to $ 0.8, 1.2, 1.6$ in the bottom row. 
At larger $|\hat{s}_1|$, mass elements are more likely to spread into vortical regions,
although part of the spreading is due to our finite bin size of $\hat{s}_1$. 
}
\end{figure*}

In Fig.\ (\ref{fig:PDF_za_theory}), we show the two-dimensional probability 
distribution of $\ts_3-\ts_2$ at two different $\ts_1$'s.
For the dispersion, we use $\sigma=0.1$ to represent an initial epoch in the first row, 
and $\sigma=1$ for a later time in the second. 
Similar to previous figures, the dashed lines illustrate the boundaries between 
classifications. 
Since initial conditions with the same $\ts_1(\tau_{init})$ but different $\ts_2(\tau_{init})$ 
and $\ts_3(\tau_{init})$ do not necessarily evolve to the same $\ts_1(\tau_{final})$, 
a comparison of the two-dimensional probability distribution at given $\ts_1$ 
between two epochs is by no means rigorous as they do not consist of the same set of samples. 
Therefore, the values of $\ts_1$ for plotting the figure are chosen more or 
less arbitrarily. 
Nevertheless, one still notices that the asymmetry increases with $\sigma$.
This statement remains valid and is even more
evident for $\ts_1=0$ since initially $\ts_{2,3}$ are symmetrically distributed. 
However, the skewed distribution in the lower panels does not necessarily mean the
probability of a Lagrangian fluid element belonging to a particular morphology 
changes with time significantly in the ZA, because $\ts_1$ as well as the morphology boundaries 
evolve together and the sample points contributing to the last panel at $\ts_1=0$ come from 
various initial $\ts_1(\tau_{init})$.

\subsection{Nonlinear Evolution beyond Zel'dovich Approximation}
Even without the shell-crossing, the gravitational nonlinearity of the system 
(Eq. \ref{eqn:Poisson_eq}-\ref{eqn:Aij_evl_full}) already makes the analytical study of
the evolution of velocity gradient invariants very complicated. 
One possibility is to investigate the Lagrangian evolution of all relevant 
quantities at least numerically, similar to Eq. (\ref{eqn:inv_za_evolution}) in ZA, 
but including density $\delta$, velocity gradient $A_{ij}$ as well as the
tidal tensor $\varepsilon_{ij}=\varpi_{ij}- \varpi\delta_{ij}/3$, which however, does 
not exist in Newtonian cosmology due to the nonlocality of the theory.
In the following, we will only concentrate on the numerical measurement from the 
simulation.

In Fig.\ (\ref{fig:pdf_2d_simulation}), we show the two-dimensional probability 
distribution of the invariants from the $N$-body simulations 
`$Sing100$'. Instead of measuring the invariants themselves, we normalize the velocity 
gradient tensor
\begin{eqnarray}
\label{eqn:Aij_rescale}
\hat{A}_{ij} = \frac{ A_{ij}}{\sqrt{A_{mn}A_{mn}} }
\end{eqnarray} 
first, given that any rescaling of $A_{ij}$ with a positive constant would not 
change the boundaries between kinematic classes.
After this normalization, the invariants are  confined within the 
cubic region $\hat{s}_1 \in [-\sqrt{3}, \sqrt{3}]$, $\hat{s}_2 \in
[ -1/2, 1 ]$ and $\hat{s}_3 \in [-\sqrt{3}/9, \sqrt{3}/9]$. 
From top to bottom, left to right, we plot the distribution at 
$\hat{s}_1=(-1.6,-1.2,-0.8, -0.4, 0, 0.4, 0.8, 1.2, 1.6)$. 
Before further preceding, one should realize that
a direct comparison between Fig.\ (\ref{fig:PDF_za_theory}) and 
(\ref{fig:pdf_2d_simulation}) could be misleading, because here we are measuring the 
volume-weighted probability distribution in Eulerian space, whereas 
Fig.\ (\ref{fig:PDF_za_theory}) is mass-weighted. 
A consistent comparison of various theoretical models as well as simulations
in both Eulerian and Lagrangian space is indeed desired, but will not be the main topic 
of this paper.

Note that the whole dataset is divided into nine thick $\hat{s}_1$ bins, while the dashed 
boundaries in each panel correspond to the median value of $\hat{s}_1$ in each bin.
In panels with medium $\hat{s}_1$ where the shape of the distribution is sharp and 
clear, one would expect most of the offset is due to this finite bin size effect. 
This includes at least from the third to the sixth panel in the figure, 
because the sample points at these $\hat{s}_1$ bins are much more than the other.
The fraction of data in all nine panels are $\{ 0.003, 0.009, 0.38, 0.31, 0.15, 
0.06,  0.041, 0.029,  0.018\}$ respectively.
For the first two panels at $\hat{s}_1=-1.6$ and $-1.2$, however, theoretical argument 
may suggest the same since it corresponds to the outflowing void region where 
gravitational nonlinearity is less significant.
In the bottom panels with large positive $\hat{s}_1$, in which regions the fluid
elements collapse fast enough, the PDFs are much flatter, and there is much spreading, especially in the last two panels.
This could be attributed to both the nonlinear evolution as well as the effect of
shell-crossing.

\begin{figure}[htp]
\begin{center}
\includegraphics[width=0.48\textwidth]{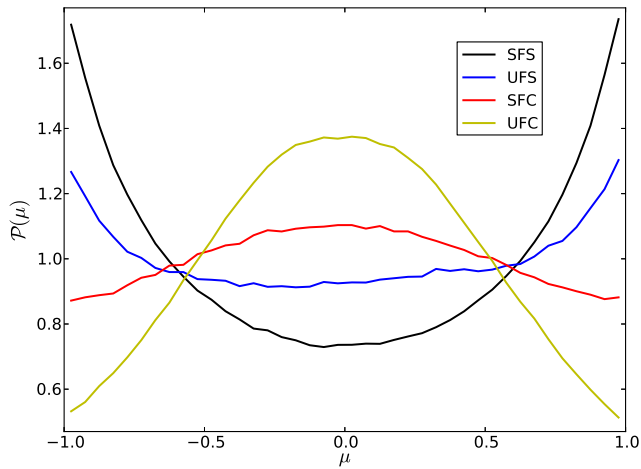}
\end{center}
\caption{ \label{fig:consine_vel_vort} 
PDF of the cosine of the angle between velocity and vorticity as measured in 
all $64$ realizations of MIP simulation. 
The alignment is strongest for categories `$\mathcal{SFS}$' and `$\mathcal{UFC}$', 
which trace filamentary structure. 
}
\end{figure}

\section{Shell-crossing and Emergence of Vorticity}

In the multi-streaming regime, vorticity is often generated.  Since velocity and density perturbations are strongly coupled,
it is reasonable to expect that rotational degrees of freedom correlate 
with large-scale structures, as in the potential flow. 
Indeed, from the N-body simulation, we do observe this associations not only via 
the distinct spatial distribution of various vortical flows, but also the alignment 
between the vorticity and the cosmic web. 
The theory, as will be seen later in this section, is complicated
and highly non-trivial quantitatively. 
One obstacle is to incorporate the shell-crossing in a closed dynamical fluid system. 
To do so, we start from the fundamental Vlasov-Poisson system, and then derive 
evolution equations of invariants including phase space information of multi-streaming. 
However, instead of solving the dynamical system, we alternatively try to propose a 
statistical description which relates to the internal structure of the invariants
space.

\subsection{The Spatial and Orientation Distribution of Vorticity}

\begin{figure*}[htp]
\begin{center}
\includegraphics[width=1.\textwidth]{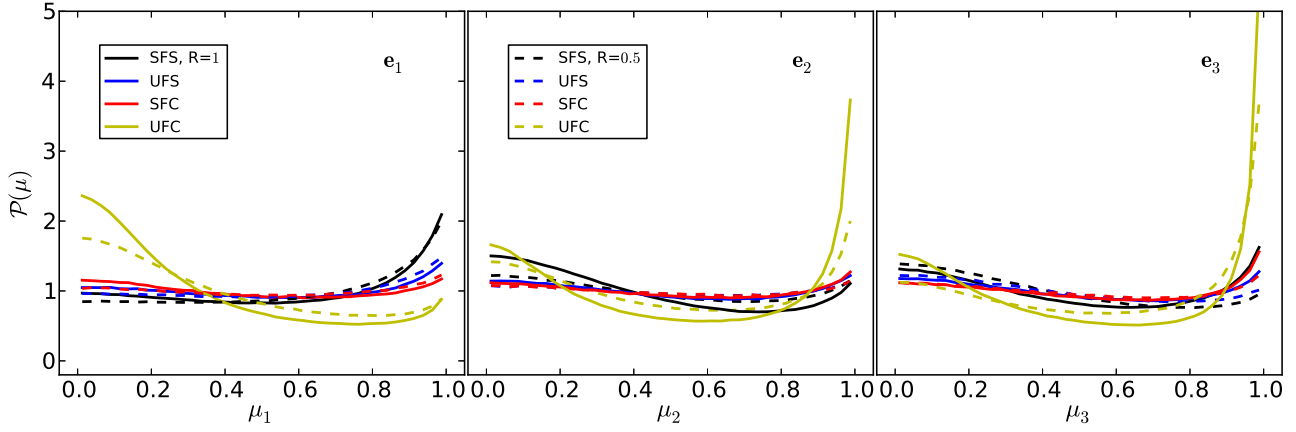}
\end{center}
\caption{ \label{fig:vort_dhessian} The alignment between the vorticity and the eigenvectors
$e_i$ of the density Hessian matrix for various rotational categories. The density 
field is smoothed with a Gaussian filter with smoothing length $R=1 \Mpc/h$ and 
$0.5\Mpc/h$, as shown in solid and dashed lines, respectively. Similar to Fig.\ 
(\ref{fig:consine_vel_vort}), the alignment is strongest for categories `$\mathcal{SFS}$' 
and `$\mathcal{UFC}$', which trace filamentary structure.  
}
\end{figure*}

The distinct spatial distributions revealed in both potential and rotational flows
suggest that the emergence of vorticity could in principle relate to the cosmic web.
Unlike in irrotational flow, the spatial distribution of all categories shown in 
Fig.\ (\ref{fig:classification_vortical}) are visually correlated with filaments, walls, and knots, where
shell-crossing takes place.
This is especially clear for the categories `$\mathcal{SFS}$' and `$\mathcal{UFC}$' with
smaller $n_{ct}$, i.e. the last row in the figure, when almost all visible filaments have been painted by those labels.
Together with flow trajectories illustrated in Fig.\ (\ref{fig:flow_category}), the 
spatial distribution of some categories also seems to be consistent with physical 
intuition. 
For example, within knots, the spatial contraction and the energy 
transfer from direct infall to orbiting leads to the inward winding and 
compression along the radial direction, i.e. the `$\mathcal{SFC}$' category.
And the type of vorticity generated along filaments funneling matter to knots should involve inward spiraling and stretching. 
However, one also notices that this agreement is not as large as one would naively 
expected. On the one hand, this reveals the complexity of the problem; on the other hand,
it might suggest a statistical/stochastic view of the vorticity generation.

This interpretation can be further verified by examining the alignment
between the direction of the vorticity, the velocity and the cosmic web structure. 
In Fig. (\ref{fig:consine_vel_vort}), we first present the distribution of the angle
between vorticity and velocity for various categories.
Two of them, which interestingly mainly trace the filamentary structures, show 
the most significant correlation signal between the direction of these two vectors, 
but in an opposite way.
In `$\mathcal{SFS}$,' the velocity tends to be aligned with vorticity, and
in type `$\mathcal{UFC}$,' the velocity tends to be perpendicular.
In general, although not as obviously as in previous examples, mass elements seem to flow along the vorticity direction if it is stretching, 
and tends to flow perpendicularly when compressing along the direction of vorticity.

Fig.\ \ref{fig:vort_dhessian} shows the alignment of the vorticity with cosmic-web structure.  We display the angle distribution between the vorticity and eigenvectors 
of the density Hessian matrix $\nabla_i \nabla_j \rho$, for 
different vortical categories. 
Here $e_i$ denotes the $i$th eigenvector, with eigenvalue $\lambda_i$, 
where $\lambda_i$ are sorted according to the absolute value, $|\lambda_1|\le 
|\lambda_2|\le |\lambda_3|$. Therefore, $e_1$ corresponds to the direction with
slowest changing rate of density distribution, and $e_3$ the fastest. 
In filaments, $e_1$ is aligned with the filament's axis.
We smoothed the density field with a Gaussian filter at
smoothing length $R=0.5 \Mpc/h$ and $1\Mpc/h$. 
Similar to the situation in velocity-vorticity alignment, 
the correlation is strongest for categories `$\mathcal{SFS}$' and 
`$\mathcal{UFC}$', which trace the filaments.
Particularly, type `$\mathcal{SFS}$' is more aligned with the 
direction of smallest eigenvalues $|\lambda_1|$.
On the other hand, one could also see the distinctive alignment of `$\mathcal{UFC}$', where the vorticity is perpendicular to $|\lambda_1|$, and parallel with 
$|\lambda_2|$ or $|\lambda_3|$. 
We also notice the smoothing variations that has been encountered by \cite{CP12}
and \cite{AC13}.

Fig.\ (\ref{fig:consine_vel_vort}) together with Fig.\ (\ref{fig:vort_dhessian}) reinforce our
previous physical interpretation. For instance, they show that the vorticity
direction of `$\mathcal{SFS}$' type aligns with both velocity as well as the filament
directions. 
Given its schematic flow trajectory from Fig.\ (\ref{fig:flow_category}), this
is consistent with the physical picture of the accretion along filament
with additional rotation around the main path \citep{PP11,CP12,LP13}, which is expected
to be generated during the formation of filamentary structure \citep{ZA70,BKP96,PP11}.
For category `$\mathcal{UFC}$', since the vorticity direction is perpendicular to both
filament and velocity, the major component of the velocity is still along the filament. 
This suggests that vorticity of this type is related to halo spin formed by major mergers
along the filaments \citep{CP12}.
Therefore as demonstrated above, shell-crossing and generation of vorticity are physically 
rich processes. In the invariants space, this means certain types of trajectories 
exist transitioning from potential flow to a particular rotational flow region, 
which as will be seen later is the main focus of this section.

\subsection{Emergence of Vorticity: Dynamical View}
Before discussing our statistical view of vorticity generation, let us go back to 
examine the dynamical system first.  To proceed, one has to restore the extra source 
contribution of the velocity dispersion in the Euler equation (\ref{eqn:euler_eq}), 
because the term $\varpi_{ij}$ in Eq.\ (\ref{eqn:Aij_evl_full}) is symmetric, therefore 
initial irrotational flow will not be able to generate vorticity.
The standard approach is to start from the more fundamental Vlasov equation of 
one-particle phase-space density $f(\vx,\vp,\tau)$ \citep{P80,BCGS02}, 
\begin{eqnarray}
\label{eqn:vlasov_eqn}
\frac{\partial f}{\partial \tau} + \frac{\vp}{am} \cdot \frac{\partial f}{\partial 
\vx} - a m \nabla \Phi \cdot \frac{\partial f}{\partial \vp} = 0, 
\end{eqnarray}
and then take moments of the velocity,  given the fluid quantities defined as
\begin{eqnarray}
 \rho &= & \int d^3\vp~ f(\vx, \vp,\tau), \quad
 u_i  = \int d^3\vp ~ f(\vx, \vp, \tau) ~ \frac{p_i}{am},  \nonumber \\
 \varsigma_{ij} &=& \int d^3\vp ~f(\vx, \vp, \tau) ~ \left(\frac{p_i}{am} - u_i \right)
\left ( \frac{p_j}{am} - u_j \right), 
\end{eqnarray}
Here $\vx$ and $\vp=am\vu$ are the comoving positions and momenta of particles, with 
$m$ the mass of particle and $\vu$ the peculiar velocity. 
And $\varsigma_{ij}$ is the velocity dispersion, and has been ignored in the previous 
dust model. Eventually one recovers the Euler equation with an extra term,
\begin{eqnarray}
\label{eqn:euler_eq_disp}
\frac{\partial \vu}{\partial \tau} + \mH(\tau) \vu(\vx, \tau) + 
\vu(\vx, \tau)\cdot \nabla \vu (\vx, \tau) = \quad \qquad \nonumber \\
-\nabla \Phi(\vx,\tau) -\frac{1}{\rho}\nabla_j (\rho\varsigma_{ij}). \qquad
\end{eqnarray}
The gradient tensor $A_{ij}$ reads as
\begin{eqnarray}
\label{eqn:Aij_evl_vort}
\frac{d A_{ij}}{d \tau} + \mH(\tau) A_{ij} + A_{ik} A_{kj} = \varpi_{ij} + \zeta_{ij},
\end{eqnarray}
and here we define the `dispersion tensor' $\zeta_{ij}$ as
\begin{eqnarray}
\zeta_{ij}= -\nabla_j \left [\frac{1}{\rho} \nabla_k\left (\rho \varsigma_{ik}\right) 
\right].
\end{eqnarray}
Replacing $\varpi_{ij}$ and its derivatives with $\varpi_{ij}+\zeta_{ij}$
in Eq.\ (\ref{eqn:inv_dyn_full}), one obtains dynamical equations of 
invariants with extra source terms,
\begin{eqnarray}
\frac{d }{d\tau} s_1 + \mH(\tau)s_1 -  s_1^2 +  2 s_2  &=& -\varpi - \zeta\
 \nonumber \\
\frac{d}{d\tau} s_2  + 2\mH (\tau)s_2  - s_1 s_2  + 3 s_3 & =& - s_1 
(\varpi +\zeta) - \varpi_{A} -\zeta_{A}  \nonumber \\
\frac{d}{d\tau} s_3  + 3\mH(\tau) s_3 -s_1 s_3 &=&  -s_2 (\varpi + \zeta) 
- s_1 (\varpi_{A} \nonumber \\
&&  + \zeta_A)  - \varpi_{A^2} -\zeta_{A^2}. \quad
\end{eqnarray}
However, the system has to be closed, i.e. the infinite 
hierarchy series needs to be truncated \citep{PS09}. A handful of suggestions have 
been proposed \citep{PS09,AD00,BD05}. In general, one desires to know the
full phase space information from Eq. (\ref{eqn:vlasov_eqn}).

\subsection{Emergence of Vorticity: Statistical View}

Instead of explicitly solving the dynamical system discussed previously, 
as will be seen later, it is also possible to establish a statistical 
description for generation of vorticity in the invariants space. 
In the multi-streaming region, one defines the bulk velocity which projected 
from the phase space as the density-weighted average among all streams 
\begin{eqnarray}
\label{eqn:vu_multi}
\vu(\vx)= \frac{1}{\rho(\vx)} \sum_s \rho(\vq_s) \vu(\vq_s) ,
\end{eqnarray}
where $\rho(\vq_s)$ and $\vu(\vq_s)$ are density and velocity of each stream, 
and $\rho(\vx)=\sum_s \rho(\vq_s)$ is the Euerlian density. 
Then by taking the gradient of Eq.\ (\ref{eqn:vu_multi}), one notices two 
separate contributions to the averaged gradient tensor $A_{ij}$: 
first is the density average of $A^s_{ij}$ among all streams
\begin{eqnarray}
\label{eqn:stat_Aij_sym}
 \frac{1}{\rho(\vx)} \sum_s\left[  \rho(\vq_s) A^s_{ij}(\vq_s) \right], 
\end{eqnarray}
and the other is the coupling between the density gradient and velocity  
arising purely due to the projection of a multi-valued field from the phase 
space \citep{PB99,HA14}, 
\begin{eqnarray}
\label{eqn:stat_Aij_wij}
\frac{1}{\rho(\vx)}\biggl[ \sum_s  u_i(\vq_s) \frac{ \partial } 
{\partial x_j} \rho(\vq_s) 
- u_i(\vx)  \frac{ \partial } {\partial x_j}\rho(\vx)  \biggr]. \qquad
\end{eqnarray}

In the cold-dark-matter scenario, matter is assumed to reside on a 
three dimensional thin sheet in the six dimensional phase space.
The generalized Kelvin's circulation theorem then ensures that the average vorticity
will remain zero within any circle in phase space \citep{L67}, and it is also
zero for all closed loops in three-dimensional position space which are non-intersecting when projected from six dimensions. 
However, in the multi-streaming region after shell-crossing, a simple circle in 
phase space could be projected down as interacting `$8$'-like loops, with nonzero 
vorticity for each of them in three-dimensional position space.
Yet, depending on the scale of interest, at any epoch, one is free to choose 
loops have not been deformed large enough to generate any net effect except near
the vicinity of singular points, which might become non-negligible in the deeply
nonlinear regime, for example in viralized halos.

For simplicity, we would like to assume the velocity of each stream is
still potential after shell-crossing, so the first contribution 
(Eq. \ref{eqn:stat_Aij_sym}) of $A_{ij}$ is symmetric.
Then, rotation arises from the coupling between density
gradient and the velocity of streams (Eq. \ref{eqn:stat_Aij_wij}).
Since various morphological structures are distinct in the density gradient as
well as their coupling with the velocity field, this contribution  
depends on the environment.  This could be responsible for the 
different spatial and orientation distribution of various vortical categories.

\subsubsection{Shell-crossing in the Invariants Space}
Given the expression of $A_{ij}$ in Eq.\ (\ref{eqn:stat_Aij_sym}) and
(\ref{eqn:stat_Aij_wij}), it is possible to set up a statistical description of the 
 vorticity emergence in the invariants space. 
At arbitrary Eulerian position $\vx$, we assume there are many streams, labeled 
from $1$ to $\alpha$, flowing potentially, and we denote the invariants as
\begin{eqnarray}
^{1\cdots\alpha}s^{(s)}_i  = 
\{ ^{1}s^{(s)}_{i}, ~ ^{2}s^{(s)}_i, ~ \cdots  ~ ^{\alpha}s^{(s)}_i \},
~ i=(1,2,3) \qquad
\end{eqnarray}
where $^{m}s^{(s)}_i$ are potential invariants for $m$th stream, and they have
the probability distribution
\begin{eqnarray}
\mP\left( ^{1\cdots\alpha}s^{(s)}_{i}\right).     
\end{eqnarray}
As seen in previous sections, this distribution characterizes the morphological
information of the cosmic web.  
After a short yet finite amount of time $\Delta \tau$, the winding of the dark matter
sheet in the phase space becomes large enough to form a multi-streaming region. 
These $\alpha$ streams encounter and couple with each other to generate final 
invariants $s_i$ through Eqs.\ (\ref{eqn:stat_Aij_sym}) and (\ref{eqn:stat_Aij_wij}),
with distribution functions $\mP(s_i)$.
This shell-crossing process can then be described by the conditional
probability distribution of final $s_i$'s given initial invariants 
$ ^1s^{(s)}_{i} \cdots ~^{\alpha}s^{(s)}_{i}$
\begin{eqnarray}
\mP(s_i| ~  ^{1\cdots\alpha}s^{(s)}_{i}). 
\end{eqnarray}
From Eq.\ (\ref{eqn:stat_Aij_wij}), these distributions also encode information about 
cosmic-web morphologies, since velocity and density gradients couple 
differently depending on the environment.
Here we explicitly write down the time lapse $\Delta \tau$ to simplify the
description, since in reality, multi-value regions emerge gradually, from
three-flow regions to more complicated cases. 
In this situation, if shell crossings occur more than $m$ times during 
$\Delta \tau$, what happens is essentially a random walk in invariants space, where the 
final distribution $\mP_{\Delta \tau}(s_i| ~  ^{1\cdots\alpha}s^{(s)}_{i} ) $ 
is a series multiplication of separate probability functions
\begin{eqnarray}
 \mP_{\Delta \tau}(s_i| ~  ^{1\cdots\alpha}s^{(s)}_{i} ) = \prod_{m}
\mP_{\tau_m}(  ^{ 1\cdots\alpha}s_i (m)| ~  ^{ 1\cdots\alpha}s_i (m-1) ). \nonumber\\
\end{eqnarray}
Here we assume at each step, only a subset $\tau_m(1\cdots\alpha)$ was involved in
the process.

\subsubsection{Boundary Crossing and the Internal Structure of Invariants Space}
To further investigate the validity of this description, we explore the probability
distribution $\mP(s_i|^{1\cdots\alpha}s^{(s)}_i)$ numerically,
and show the result in Table (\ref{tab:vort}).
At an arbitrary location where shell-crossing occurs, we assume that a random number of 
streams overlap.  From Eq.\ (\ref{eqn:stat_Aij_sym}) and Eq. \
(\ref{eqn:stat_Aij_wij}), both the symmetric and anti-symmetric contributions are
environmentally dependent, so we first generate samples of $A^s_{ij}$ for 
various kinematical categories from Zel'dovich approximation, and identify these 
irrotational kinematical types as morphology structures.  That is, we
consider `$\mathcal{SN/SN/SN}$' as halo, `$\mathcal{SN/S/S}$' as filament and
`$\mathcal{UN/S/S}$' as wall structures.  
Then based on their morphological classes, we generate velocity as well as the 
density gradient for all flow elements, which roughly resemble the physical process
near each morphological structure. For example, both the velocity vectors and density 
gradients are sampled spherically for halos. The velocities are assumed to flow close
to a line for filaments, and the density gradients around the radial direction of 
a cylinder. And near the wall, both velocities and density gradients are chosen close
to a given direction.
With $u_i$, $\nabla_i \rho$, and $A^{s}_{ij}$ for each stream, we are then able to 
construct the full velocity gradient tensor $A_{ij}$ via Eqs.\ (\ref{eqn:stat_Aij_sym})
and (\ref{eqn:stat_Aij_wij}).

\begin{table*}[ht]
\begin{center}
\caption{ \label{tab:vort}  Kinematic morphologies generated by the toy model of
 stream-crossing}
\begin{tabular}{|c|c|c|c|c|c|}
    \hline
    & $\mathcal{SN/SN/SN}$ ({\it halo}) & $\mathcal{SN/S/S}$ ({\it filament}) &
      $\mathcal{UN/S/S}$ ({\it wall }) & $\mathcal{UN/UN/UN}$ ({\it void }) & 
      $total~(after)$ \\
    \hline
    $\mathcal{SN/SN/SN}$&$0.3~(10.1)$&$0.5~(1.6)$  &$0.3~(0.5)$  & $-$ &   $1.1 $\\
    $\mathcal{SN/S/S}$ & $1.1~(34.3)$&$10.5~(37.0)$&$11.3~(22.5)$& $-$ &   $22.9$\\
    $\mathcal{UN/S/S}$ & $0.2~(6.0)$ &$4.3~(15.1)$ &$25.8~(51.3)$& $-$ &   $30.3$\\
    $\mathcal{UN/UN/UN}$&$0.0~(0.0)$ &$0.1~(0.2)$  &$3.6~(7.1)$  &$18.2~(100)$&$21.8$\\
    $\mathcal{SFS}$ &    $0.4~(13.2)$&$5.3~(18.5)$ &$2.6~(5.2)$  & $-$ &   $8.3 $\\
    $\mathcal{UFS}$ &    $0.0~(0.4)$ &$0.6~(2.1)$  &$2.5~(5.0)$  & $-$ &   $3.1 $\\
    $\mathcal{SFC}$ &    $0.8~(26.5)$&$2.3~(8.0)$  &$0.6~(1.1)$  & $-$ &   $3.7 $\\
    $\mathcal{UFC}$ &    $0.3~(9.5)$ &$4.9~(17.4)$ &$3.6~(7.1)$  & $-$ &   $8.8 $\\
    \hline
    $total~(before)$ & $3.2~(100)$ &  $28.4~(100)$ &  $50.2~(100)$  & $18.2~(100)$ & $100$\\
    \hline
\end{tabular}   
\\[8pt]
Each column shows the percentage of flow morphologies generated from particular type 
displayed at the top, numbers in parentheses give the normalized fraction within
every column, i.e. the probability of generating other categories from that
morphology.
The last row/column is the marginal sum of each column/row. 
Therefore, the last row gives the total fraction of each morphologies before 
the crossing, and the last column gives total percentage 
after crossing. To compare with the spatial distributions in Fig.\
 (\ref{fig:classification_vortical}), the pre-crossing fractions, i.e. the last
row, are volume-weighted in Eulerian space at redshift $z=2$.
All numbers are displayed in percentage (\%), and have been rounded for display purpose.
The dashes in the column of void means our model does not assume any stream-crossing 
in void region, and therefore they are set to be zero.
\end{center}
\end{table*}

We also list the fractions of all kinematical 
types generated by our numerical shell-crossing model in Table (\ref{tab:vort}).
The total number of samples we generated is the roughly the order of $256^3$.
Each column corresponds to a particular morphological structure we started from, 
which from left to right are halo `$\mathcal{SN/SN/SN}$', filament 
`$\mathcal{SN/S/S}$', wall `$\mathcal{UN/S/S}$' and void `$\mathcal{UN/UN/UN}$'. 
In our probably over-simplified toy model, stream-crossing happens for streams with 
the same kinematical class. After that, however, all categories can be generated,
including both potential and vortical classes, as shown in different rows.
In the table, we list in each column the percentage of flow morphologies generated from the particular type displayed at the top, numbers in parentheses give the normalized 
probability within every column. The last row gives the total volume weighted fraction of 
each morphologies before the crossing, and the last column gives total percentage
after the crossing.
 
From the table, physically implausible situations, like voids generated in
highly multi-streaming filaments and halos, are produced only in very tiny amount.
Although a substantial fraction of potential flows are produced, we are more 
interested in the vortical ones.
From the table, there are higher probabilities to generate vorticity with
class `$\mathcal{SFC}$' around halos ($26.5\%$ compared with $13.2\%$ of 
`$\mathcal{SFS}$' and $9.5\%$ of `$\mathcal{UFC}$'); 
type `$\mathcal{SFS}$' ($18.5\%$) as well as `$\mathcal{UFC}$' ($17.4\%$) 
near filaments; and a similar fraction of almost all vortical types around walls.
Given the simplicity of our model, the exact value in Table (\ref{tab:vort}) should 
not be taken too seriously. 
However, from the fifth row of the table, among $8.3\%$ of vortical category 
`$\mathcal{SFS}$' being generated from stream crossing, more than sixty percents of them 
are formed near filaments `$\mathcal{SN/S/S}$'. And this is also true for category
`$\mathcal{UFC}$' ($4.9\%$ compares with $8.8\%$ in total). 
From Fig.\ (\ref{fig:classification_vortical}), this is exactly what one found from 
the filamentary spatial distribution of these two categories.
For type `$\mathcal{SFC}$', however, due to small fraction of type `$\mathcal{SN/SN/SN}$'
initially, the dominant contribution comes from filaments. This might also be 
physically plausible since the spatial distribution of this type tends to extend 
a little from central halos to the filamentary structures.
Finally, both our model and Fig.\ (\ref{fig:classification_vortical})  suggest type
`$\mathcal{UFC}$' is more likely to be generated in walls ($2.5 \%$ compares with 
$3.1\%$ in total). 

Even without any detailed information about the conditional distribution of
the shell-crossing $\mP(s_i|^{1\dots\alpha}s^{(s)}_i)$, it is not hard to understand
the result, since from Fig.\ (\ref{fig:flow_category}), one discovers that the region
 of irrotational flow that follows the filament structures (`$\mathcal{SN/S/S}$') is 
just adjacent to the `$\mathcal{SFS}$' vortical region in invariants space, which
 also is associated with filaments. Similarly, the triangle-like region of 
halos `$\mathcal{SN/SN/SN}$' resides just inside the `$\mathcal{SFC}$' regime.
Physically, this is also consistent with the formation of cosmic web structures.
Consider mass elements flowing around filaments potentially. In 
the canonical form of the tensor $A_{ij}$, there are three real eigenvalues $\lambda_i$. 
At some point, shell-crossing occurs, and from Zel'dovich's structure-formation theory, the filaments are formed from the second collapse, after walls. 
The trajectory in position space will produce a spiral in the plane perpendicular to the filament.
In the idealized case where $\lambda_1\sim\lambda_2<0$ for any of two eigenvalues 
initially, and only small amount $\epsilon$ has been changed to the real part of 
$\lambda_1$, one obtains the canonical form 
\begin{eqnarray}
\label{eqn:Aij_canonical_asym_shellcrossing}
\vA =  \left( \begin{array}{ccc}
\lambda_1 +\epsilon & -b & \quad   \\
b  & \lambda_1 +\epsilon & \quad  \\
 \quad & \quad & \lambda_3   \\
\end{array}   \right ).
\end{eqnarray}
From Eq.\ (\ref{eqn:vort_si_canonical}), in this case, $s_2$ will be
enlarged and $s_3$ will be smaller, giving a shift in the upper-left 
direction in the invariants space. So, just around filaments, category 
`$\mathcal{SFS}$' will be generated, and the spatial association with the large-scale
structure is preserved after shell-crossing. 
Furthermore, this physical picture also suggests that the vorticity direction is aligned
with filaments, consistent with our measurement in simulation for this category.
The situation around halos is even simpler, since from Fig. (\ref{fig:flow_category}),
category `$\mathcal{SN/SN/SN}$' is almost entirely surrounded by the `$\mathcal{SFC}$'
type in the invariants space. 
And therefore it gives much higher probability to generate this type of rotational flow
around halos.

\section{Conclusion and Discussions}
In this paper, we concentrated on the velocity gradient tensor $A_{ij}$ and its rotational 
invariants defined as coefficients of the characteristic equation of $A_{ij}$.
They enable one to classify all possible flow patterns including both potential and 
rotational flows. 
For irrotational flow, our invariants-based categorization is equivalent to
the cosmic web finder `V-web' \citep{HM12} with the threshold eigenvalue $\lambda_{th}=0$.
The PDF of density and fractional anisotropy of various flows are consistent with
their morphology types, although the visual impression of the web structure might not be as
accurate as other techniques \citep{PP09,AC10,S11a,SP11b,BSC10,FR09,HP07a,SCP09,SM05,HM12}.
To improve it, one could introduce a non-zero fine-tuned threshold $\lambda_{th}$ as done in
the `V-web' finder. In our method, this can be achieved by a simple invariants 
transformation (Eq. \ref{eqn:s_redfine_th}). 
However, from the fractional anisotropy distribution, the physical interpretation might
become difficult. 
Since all invariants are one-to-one mapped to eigenvalues, they could also be applied 
to other dynamical classification methods, e.g. invariants defined for deformation 
tensor or the Hessian matrix of gravitational potential \citep{HP07a,FR09}.

As a region develops vorticity, the invariants change continuously, which is a clear 
benefit of working with these invariants. 
By combining an ensemble of $N$-body simulations with the same large-scale modes, we 
show that various rotational flows also trace the cosmic web structure differently.
This reveals the dynamical nature of these variables and provides an alternative 
view on the emergence of vorticity. 
Therefore, an understanding of their dynamical evolution would be valuable. 
We first stepped back and concentrated on the irrotational flow, which is
important on its own as the indicator of cosmic web, 
and started from the simplest Zel'dovich approximation,  where both accurate 
solution of time evolution and the analytical formula of PDF are available. 
We also wrote down a set of dynamical equations for the invariants from the Euler 
equation. However, further investigation is difficult due to the gravitational 
nonlinearity. One possibility is to study the Lagrangian evolution of all relevant 
quantities, including the density $\delta$, velocity gradient $A_{ij}$, and the 
tidal tensor $\epsilon_{ij} = \varpi_{ij} - \varpi \delta_{ij} /3$. 
Due to the non-locality of Newtonian cosmology, the evolution equation of 
$\epsilon_{ij}$ is missing, and will only be feasible under certain approximation.
We will follow this approach in a separate paper \citep{WW14}.

Given the complications of dynamical modeling of the nonlinear gravitational evolution 
as well as the shell-crossing, the key insight came after realizing the stochastic 
nature of the multi-streaming.  The distinctive spatial distribution of different
rotational flow morphologies is very likely to be a consequence of some diffusion 
process in the invariants space. Indeed, the internal structure of the invariants space 
(Fig.\ \ref{fig:flow_category}) shows the adjacency between the potential morphology 
`$\mathcal{SN/SN/SN}$' and rotational type `$\mathcal{SFC}$' who both spatially follow 
halos, and the contiguity between type `$\mathcal{SN/S/S}$' and `$\mathcal{SFS}$' which 
all appear around filaments. 
Without any sophisticated model, we numerically investigated the PDF of invariants 
generated from a toy model of stream-crossing. 
The result in Table.\ (\ref{tab:vort}) qualitatively supports our conjecture.

However, many details of this stochastic process are still missing. For example, it 
is unclear whether a simple diffusion would be able to explain the physics associated
with halo formation. Unlike the vorticity, the angular momenta of halos can 
be formed from laminar flow by the misalignment between the shear and inertia of 
given region which encloses the material of a protogalaxy, i.e. from the tidal 
torque theory \citep{H49,P69,D70,W84,WP85}.
Compared with the fast growth $\sim D^{6.5/2}$ \citep{PS09} of vorticity after its
emergence, the halo spin grows slower since the very early stages of structure
formation. However, as suggested by \cite{L13a}, the strong alignment between the direction
of vorticity and halo angular momentum suggests another phase of the coevolution 
between them after the tidal torque becomes ineffective after turn-around when the 
lever arm is dramatically reduced.

This coevolution phase seems to be reasonable since as in our findings of two 
different alignments between vorticity and filaments (category `$\mathcal{SFS}$'
and `$\mathcal{UFC}$' respectively), recent studies also suggest two states of 
preferential orientation between halo spins and filaments \citep{HP07a,SCP09,ZY09,
CP12,AC13}. Moreover, it has been confirmed that this orientation is 
mass-dependent \citep{CP12}, i.e. low-mass halos tend to be aligned with filaments, and 
high-mass halos have spin perpendicular to the filaments \citep{AC07,HC07b,PSP08,CP12}.
\cite{CP12} proposed an interpretation that more massive halos acquire their spins 
that perpendicular to the filaments via merger process along filament. If this is the
case, the trajectories of a mass element in the invariants space might be much
more complicated.
Further investigations of how such process would manifest itself in the invariants 
space would be interesting.

\acknowledgments
We thank Charles Meneveau, Roman Scoccimarro and Francis Bernardeau for inspiring 
discussions. XW, AZ and GLE are grateful for support by National Science Foundation’s CDI-II program, project CMMI-0941530.
MAAC and MCN are supported by a New Frontiers in Astronomy and Cosmology grant from the Sir John Templeton Foundation.

\appendix

\newcommand{\appsection}[1]{\let\oldthesection\thesection
\renewcommand{\thesection}{Appendix \oldthesection}
\section{#1}\let\thesection\oldthesection}

\appsection{List of Classification}
Most results in this appendix are originally from \cite{CPC90}.  We quote the results here
for the purpose of completeness. Considering the solution to the characteristic equation 
\begin{eqnarray}
\det[\vA-\lambda \vI] = \lambda^3 + s_1 \lambda^2 + s_2 \lambda + s_3 = 0,
\end{eqnarray}
one can look at the coefficient space $\{ s_1, s_2, s_3\}$.
The surface $S_1$ that divides the real and complex solutions is given by
\begin{eqnarray}
\label{eqn:solution_S1_app}
 27 s_3^2 + ( 4 s_1^3 - 18 s_1 s_2) s_3 + (4 s_2 ^3 - s_1^2 s_2^2) = 0
\end{eqnarray}
Denoting the solution of the above equation as $s_{3a}$ and $s_{3b}$, assuming $s_{3a}\le s_{3b}$,
regions with real eigenvalues at fixed $s_1$ correspond to
\begin{eqnarray}
    \label{eqn:real_cond}
    s_2 \le s_1^2 /3, \qquad  s_{3a}\le s_3 \le s_{3b}.
\end{eqnarray}
When either equal sign of the second inequality holds, two of eigenvalues will be equal. 
And if the first equality also holds, one gets three real and equal eigenvalues.

For the purpose of cosmology, we are particularly interested in the following 
potential flow categories:\\
\begin{eqnarray}
Stable ~Nodes ~(\mathcal{SN/SN/SN}): & \qquad &
~~ s_1 >0, \qquad 0<s_2<s_1^2/3, \qquad \max(s_{3a}, 0)<s_3<s_{3b}, \nonumber\\
Stable ~Node/Saddle ~(\mathcal{SN/S/S}): & \qquad &    
  ~~ s_1\ge 0, \qquad s_2<s_1^2/4, \qquad s_{3a}<s_3<0, \nonumber\\
 & or & ~~ s_1 < 0, \qquad s_2<0, \qquad s_{3a}<s_3<0. \nonumber\\
Unstable ~Node/Saddle ~(\mathcal{UN/S/S}): & \qquad   & 
 ~~ s_1\le 0, \qquad s_2<s_1^2/4, \qquad 0<s_3<s_{3b}, \nonumber \\
&or & ~~ s_1 > 0, \qquad s_2<0, \qquad 0<s_3<s_{3b}. \nonumber \\
Unstable ~Nodes ~(\mathcal{UN/UN/UN}): & \qquad &
~~ s_1<0, \qquad  0<s_2<s_1^2/3, \qquad s_{3a}<s_3<\min(s_{3b},0),  
\end{eqnarray}
where we list in the order of halo, filament, wall and void. 
And for vortical flow, we have
\begin{eqnarray}
Stable ~Focal ~Stretching ~ (\mathcal{SFS}): & \qquad &
    ~~ s_1\ge 0, \qquad s_2>s_1^2/4, \qquad s_3 <0, \nonumber\\
& or & ~~   s_1 \ge 0,   \qquad s_2<s_1^2/4, \qquad s_3 <s_{sa}, \nonumber \\
& or & ~~   s_1 < 0,     \qquad s_2>0, \qquad s_3 <s_1 s_2, \nonumber \\
& or & ~~   s_1 < 0,     \qquad s_2<0, \qquad s_3 <s_{sa},  \nonumber\\
Stable ~ Focal~ Compressing ~ (\mathcal{SFC}): & \qquad &
   ~~ s_1 > 0, \qquad s_2>s_1^2/3, \qquad 0<s_3 <s_1 s_2, \nonumber\\
& or & ~~   s_1 > 0,   \qquad s_1^2/4<s_2<s_1^2/3, \qquad 0<s_3 <s_{sa}, \nonumber \\
& or & ~~   s_1 > 0,     \qquad 0<s_2\le s_1^2/3, \qquad s_{3b}<s_3 <s_1 s_2 \nonumber\\
Unstable ~ Focal~ Stretching ~(\mathcal{UFS}): & \qquad &
  ~~ s_1 < 0, \qquad s_2>s_1^2/3, \qquad s_1 s_2<s_3 <0 , \nonumber\\
 & or & ~~   s_1 < 0,  \qquad s_1^2/4<s_2<s_1^2/3, \qquad s_{3b}<s_3 <0, \nonumber \\
 & or & ~~   s_1 < 0,    \qquad 0<s_2\le s_1^2/3, \qquad s_1 s_2<s_3 < s_{3a} \nonumber\\
Unstable ~ Focal~ Compressing ~(\mathcal{UFC}): & \qquad &
~~ s_1\le 0, \qquad s_2>s_1^2/4, \qquad s_3 >0, \nonumber\\
& or & ~~   s_1 \le 0,   \qquad s_2<s_1^2/4, \qquad s_3 >s_{sb}, \nonumber \\
& or & ~~   s_1 > 0,     \qquad s_2>0, \qquad s_3 >s_1 s_2, \nonumber \\
& or & ~~   s_1 > 0,     \qquad s_2<0, \qquad s_3 >s_{sb},  \nonumber \\
\end{eqnarray}
For a complete classification, please see the Appendix in \cite{CPC90}.

\appsection{Vorticity and Vortical Flow}
\label{app_sec:vorticity_diff}
Furthermore, we would like to discuss the distinction between
vorticity and vortical flow defined via local trajectories in our paper. 
As a simple example, consider the velocity field
\begin{eqnarray}
\vu(\vx) = x_1 \vu_0,
\end{eqnarray}
where $\vu_0$ is a constant vector and $x_1=\vx\cdot \ve_1$ along direction $\ve_1$. The flow is entirely aligned with $\vu_0$, but the amplitude varies.
One can show that the gradient of $\vu$ is anti-symmetric,
i.e.\ the vorticity is nonzero. 
However, the invariants of this expression are consistent with non-vortical flow, 
$s_2=s_3=0$. Or more precisely, it actually belongs to the degenerate kinematical
categories, which is not a main topic in this paper. 
A more physical example is near a caustic after shell-crossing.
Among all streams at given position, there are two categories of streams, 
ordinary flow $c_1$ and  singular flow $c_2$, where 
\begin{eqnarray}
J\left[\vx(\vq_{c_2})\right ] = \left| \frac{\partial \vx (\vq_{c_2})}
{ \partial \vq_{c_2}} \right| = 0. \nonumber\\
\end{eqnarray}
Consider an Eulerian position $\vx$ near the caustic edge $\vx_0$, with corresponding 
Lagrangian position $\vq$ and $\vq_0$. 
Following \cite{PB99}, there exists a direction orthogonal to the caustic edge, denoted
as $\perp$ direction, where the Eulerian coordinate
\begin{eqnarray}
(\vx-\vx_0)_{\perp} \approx -\eta (\vq_i - \vq_0)^2_{\perp}.
\end{eqnarray}
The Jacobian for two singular flows is 
$J(\vx)\approx 2\sqrt{ \eta(\vx_0-\vx)_{\perp}}$.  
Then the Eulerian velocity $\vu(\vx)$ could be expressed as
\begin{eqnarray}
\label{eqn:vu_lag}
\vu(\vx) \approx \frac{2 \vu(\vq_0) / J(\vx) + \rho(\vq_3) \vu(\vq_3) }{2/J(\vx)+
\rho(\vq_3)}, 
\end{eqnarray}
where $\vq_3$ are ordinary flow. As shown in \cite{PB99}, the vorticity of 
Eq. (\ref{eqn:vu_lag}) is nonzero. However again, the invariants of this expression 
are consistent with non-vortical flow; $s_2 =s_3=0$.

\label{lastpage}

\end{document}